\documentclass[10pt,number,amsmath,amssymb]{elsarticle}
\biboptions{sort&compress}
\usepackage{graphicx}
\usepackage{fancyhdr}   
\usepackage{amsmath}    
\usepackage{comment}    
\usepackage{caption}
\usepackage{fancybox,shadow}
\usepackage{morefloats}

\newcommand{\veps}{\varepsilon}
\newcommand{\be}{\begin{eqnarray}}
\newcommand{\ee}{\end{eqnarray}}
\usepackage{makeidx}  

\newcommand{\zdr}{${\rm ZD}_{\rm R}$}

\usepackage{cite}
\usepackage{url}

\usepackage{amsfonts}
\usepackage{xspace}
\date{}

\title{Evolutionary Game Theory using Agent-based Methods}
\author[mmg,pa,bea]{Christoph Adami\corref{cor1}}
\ead{adami@msu.edu}

\author[cse,bea]{Jory Schossau}
\ead{jory@msu.edu}

\author[ib,cse,bea]{Arend Hintze}
\ead{hintze@msu.edu}

\cortext[cor1]{Corresponding author}

\address[mmg]{Department of Microbiology and Molecular Genetics}
\address[pa]{Department of Physics and Astronomy}
\address[cse]{Department of Computer Science and Engineering}
\address[ib]{Department of Integrative Biology}
\address[bea]{BEACON Center for the Study of Evolution in Action,\\ Michigan State University, East Lansing, Michigan, USA}

\begin{document}

\begin{abstract}
Evolutionary game theory is a successful mathematical framework geared towards understanding the selective pressures that affect the evolution of the strategies of agents engaged in interactions with potential conflicts. While a mathematical treatment of the costs and benefits of decisions can predict the optimal strategy in simple settings, more realistic settings such as finite populations, non-vanishing mutations rates, stochastic decisions, communication between agents, and spatial interactions, require agent-based methods where each agent is modeled as an individual, carries its own genes that determine its decisions, and where the evolutionary outcome can only be ascertained by evolving the population of agents forward in time. While highlighting standard mathematical results, we compare those to agent-based methods that can go beyond the limitations of equations and simulate the complexity of heterogeneous populations and an ever-changing set of interactors. We conclude that agent-based methods can predict evolutionary outcomes where purely mathematical treatments cannot tread (for example in the weak selection--strong mutation limit), but that mathematics is crucial to validate the computational simulations.  
\end{abstract}

\begin{keyword}
Evolutionary game theory \sep Agent-based modeling
\end{keyword}
\maketitle

\section*{Introduction} 
Evolutionary game theory is an application of the mathematical framework of game theory~\citep{NeumannMorgenstern1944} to the dynamics of animal conflicts (including, of course, the conflicts people engage in). In game theory, the object is to find an appropriate strategy to resolve arising conflicts, or alternatively to find the optimal sequence of decisions that leads to the highest payoff. Even though game theory has been influential in economics and finance, perhaps its most well-known application has been in the life sciences. Beginning with the seminal paper by Maynard Smith and Price~\citep{MaynardSmithPrice1973}, Evolutionary Game Theory (EGT) has burgeoned into a mainstay of mathematical and computational biology (see, e.g., these textbooks and monographs~\citep{MaynardSmith1982,Axelrod1984,Weibull1995,Dugatkin1997,HofbauerSigmund1998,Nowak2006}). 

The mathematical foundations of game theory (originally due to von Neumann and Morgenstern~\citep{NeumannMorgenstern1944} and extended by Nash~\citep{Nash1950}) experienced a revival when Maynard Smith and Price turned their attention to EGT. Maynard Smith coined the term ``Evolutionary Stable Strategy" (ESS), to mean a move (or play) that would assure the type of animal that wields it an evolutionary (that is, Darwinian) advantage over the opponent, in the sense that the ESS could never become extinct (but it may have to coexist with other strategies that are also ESS).  The concept of the ESS is related to the Nash equilibrium (the rational choice of strategy in economic games), but it is more refined: while every ESS is a Nash equilibrium, some Nash equilibria are not ESS because they are unstable fixed points of the evolutionary dynamics (see, e,g,~\citep{MaynardSmith1982,HofbauerSigmund1998}). Even Maynard Smith's concept of the ESS turned out to be limited, however, because in games with more than two strategies, some games display stable fixed points that are not strictly ESS {\em sensu} Maynard Smith~\citep{Zeeman1980,HofbauerSigmund2003}. However, these stable fixed points guarantee survival, just as an ESS does. 

All of this mathematics can be boiled down to the following intuitive idea: Suppose we are presented with a population that is composed of individuals that each perform a particular action (a ``move") that is encoded in their genes (and therefore heritable). Which of these individuals (or, more precisely, which genes) will ultimately survive, given the particular payoffs between each pair of strategies? 
If the result of the interaction between any pair of players is known in advance (usually, and conveniently, encoded in a payoff matrix), then the evolution of the population over time can be determined by solving a coupled set of differential equations (the replicator equations, see, e.g.~\citep{TaylorJonker1978,Zeeman1980,HofbauerSigmund1998}). However, those equations describe an idealized situation: perfect mixing of populations (that is, offspring are placed at an arbitrary location, not necessarily surrounded by their own kind), deterministic play (no stochasticity in the decisions), and infinite population size. In fact, in this limit, the predictions of the replicator equations are identical with the predictions of the ESS (amended by Zeeman~\citep{Zeeman1980} where appropriate).

\vskip 0.25cm
\noindent\shadowbox{
\begin{minipage}{6.3in}
\centerline{\bf Box 1: Payoff Matrices and Stability}
\mbox{}\vskip 0cm
\noindent The outcome of evolutionary games is often determined by the payoff that accumulates between two players with different strategies. In the simplest case, the strategies are unconditional and deterministic: they only depend on the genes of the player, but not on the genes (or the phenotype, i.e., the actions) of the opponent. If there are only two such plays/genes, then the payoff matrix can be rendered as a $2\times2$ matrix, where each element of the matrix is the payoff to the {\em row} player, that is, the strategy ``on the left" (rather than ``on top") of the matrix. If we label the plays `C' and `D', then the most general $2\times2$ game is defined by the matrix
\be
\bordermatrix{\mbox{} & {\rm C} & {\rm D}  \cr
                           {\rm  C}      & a  & b  \cr
                            {\rm D}      &c &  d  }\;.
\ee      
It turns out that for infinite well-mixed populations (but not for populations with spatial structure, such as games on graphs\citep{SzaboFath2007,SzaboBorsos2016}), the outcome of the game is invariant upon adding or subtracting a constant to any column of the game (this holds also for games with more than two strategies). Rendered into normal form where the diagonal of the payoff matrix vanishes, the game only depends on two constants:     
\be
\bordermatrix{\mbox{} & {\rm C} & {\rm D}  \cr
                           {\rm  C}      & 0  & a  \cr
                            {\rm D}      &b &  0  }\;. \label{game}
\ee   
It is then easy to determine that there are only four different kinds of matrices--giving rise to four different classes of games (Zeeman classes~\citep{Zeeman1980}: if a game is in a Zeeman class, then an infinitesimal change in the payoffs will not move it to a different class). For two plays, the classes are given by the relative sign of the two constants $a$ and $b$~\citep{Zeeman1980,HofbauerSigmund1998,SzaboFath2007,Adamietal2012}: they are Prisoner's Dilemma ($a<0,b>0$),
Snowdrift game ($a>0,b>0$), Anti-coordination game  ($a<0,b<0$), and Harmony ($a>0,b<0$). For games with three strategies, there are 20 different Zeeman classes (10 main classes plus the 10 sign-reversed payoff matrices), and altogether 38 different phase portraits. For four
plays, there are already $2\times 114$ classes~\citep{Zeeman1980}. 
If strategies are probabilistic rather than deterministic (an agent plays C with probability $p$), the theoretical phase portraits (a geometric representation of the set of fixed points, as well as the trajectories between them) carry over unchanged from a representation in terms of population fractions (deterministic play) to probabilities equal to the population fractions, except that some unstable fixed points will convert to stable ones for games with more than two players~\citep{Adamietal2012}.
         
\end{minipage}
}

However, real populations are never infinite. Nor are they ever perfectly well-mixed. But these are only the obvious limitations. There are yet more limitations of the mathematics of game theory of perhaps greater importance. For example, it goes without saying that decisions are not always deterministic, and can also be influenced by the memory of previous encounters. But chief among the limitations probably is this: When strategies compete in an evolutionary scenario, it is unthinkable that all {\em possible} strategies compete against each other at one precise point in time. Rather, the strategies that do compete are those that are around at this one particular period in time, and the set of strategies that compete changes over time. New strategies emerge to test their mettle with the established ones, while once-dominant strategies can be forced to extinction by a newcomer. The success of a strategy, therefore, should be determined in the context of the strategies that it is exposed to, in space as well as in time.

What is described by the scenario where all existing strategies battle it out with each other is what is otherwise known as {\em microevolution}, that is, the dynamics engendered by the change of existing allele frequencies based on their relative fitnesses in the population. Strictly (and technically) speaking, this is a zero-mutation rate approximation of evolutionary theory, and the large-scale statistics associated with this process are well described by the ``first term" of the Price equation~\citep{Price1970} (applied to fitness as the trait under selection), also known as Fisher's Fundamental Theorem~\citep{Fisher1930,Edwards1994}. 

It goes without saying that micro-evolutionary dynamics is not un-interesting. However, on a broader scale of evolutionary dynamics, we are interested in the emergence of novel alleles and traits that the population has never experienced before, how these alleles fare against the established ones, and which alleles go to extinction as a consequence of the emergence of new types or due to changed environments. In other words, what is often perceived as most interesting on an evolutionary scale is the emergence of complexity when there was none before; how evolution can give rise to refined answers to seemingly intractable morphological or metabolic problems, with often highly creative solutions. Because changing environments in particular often require novel alleles for survival (in these changed conditions), a restriction to {\em existing} variation does not do justice to the fundamental creative power of the evolutionary process. An EGT that focuses on existing variation in our view therefore does not merit the `E' in EGT. We will argue here that it takes agent-based simulation methods in a game-theoretic framework to put the `E' back into EGT, and that failing to do so can obscure many important (perhaps even {\em the} most important) aspects of the evolution of cooperation. 

Arguably, the majority of the literature in EGT is mathematical in nature, with simulations either used to validate the mathematical arguments, or else to investigate limits in which closed form solutions are unavailable. We cannot here do justice to all this literature, and often refer instead to a number of excellent textbooks or reviews (see, e.g.,~\citep{SzaboFath2007} for a comprehensive review of the literature covering populations evolving on grids or graphs). Instead, we choose here to highlight those elements of evolutionary game dynamics that not only prevent closed form solutions, but where it is not even clear how to write down the equations. We argue that agent-based simulations provide a means to move beyond mathematics~\citep{Holland2012,Adami2012}, without loss of rigor, but with a significant gain in predictive power.

We first discuss the limitations imposed by a finite population size, and conclude that this is only a limitation in conjunction with other confounding variables such as mutation rate. We then examine the role of finite mutation rates in evolutionary dynamics, and in particular examine the mathematically intractable ``weak selection-strong mutation" regime, while arguing that all realistic biological populations operate in that regime.
We then consider the impact of stochastic rather than deterministic strategies on evolutionary stability,  with applications to games with cyclic dominance as well as evolutionarily stable sets (ES sets). We subsequently consider strategies that can take advantage of information to make decisions (both deterministic and stochastic), and discuss the evolutionary stability of a subclass of communicating (that is, conditional) strategies, the so-called ``Zero-Determinant" strategies. After investigating their stability in the strong mutation regime, we look into their extension into multi-player games, and study the effect of communication in the guise of punishment in those games. Finally, we briefly survey games on graphs, for which some analytical results can be obtained, but where most of the results today come from agent-based simulations.  

\section*{Agent-based methods}
\subsection*{Limitations due to finite population size}
Finite populations make the mathematical solution of game theoretic dilemmas more difficult, but not impossible. 
In infinite populations, when using the payoff matrix (\ref{game}) the fraction of the population of type C is determined 
by the ordinary differential equation (the replicator equation)
\be
\dot x_C(t)=x_C(t)*(1-x_C(t))\left[-(b+a)x_C(t)+a\right]\;,   \label{replic}
\ee
where the density of defectors is $x_D(t)=1-x_C(t)$.
If populations are not infinite, then the replicator equation approach will not correctly predict the population outcome anymore. Indeed, the infinite population limit Eq.~(\ref{replic}) predicts two trivial fixed points (``absorbing states)" $x_C=0$ and $x_C=1$ (along with a possible mixed population state), the former being stable while the latter is unstable. At finite populations the transition from $x_C=1/N$ to zero (or from $x_C=1-1/N$ to 1) represents an irreversible transition: the extinction of the last representative of the alternative strategy.

\begin{figure}[htbp] 
   \centering
   \includegraphics[width=5in]{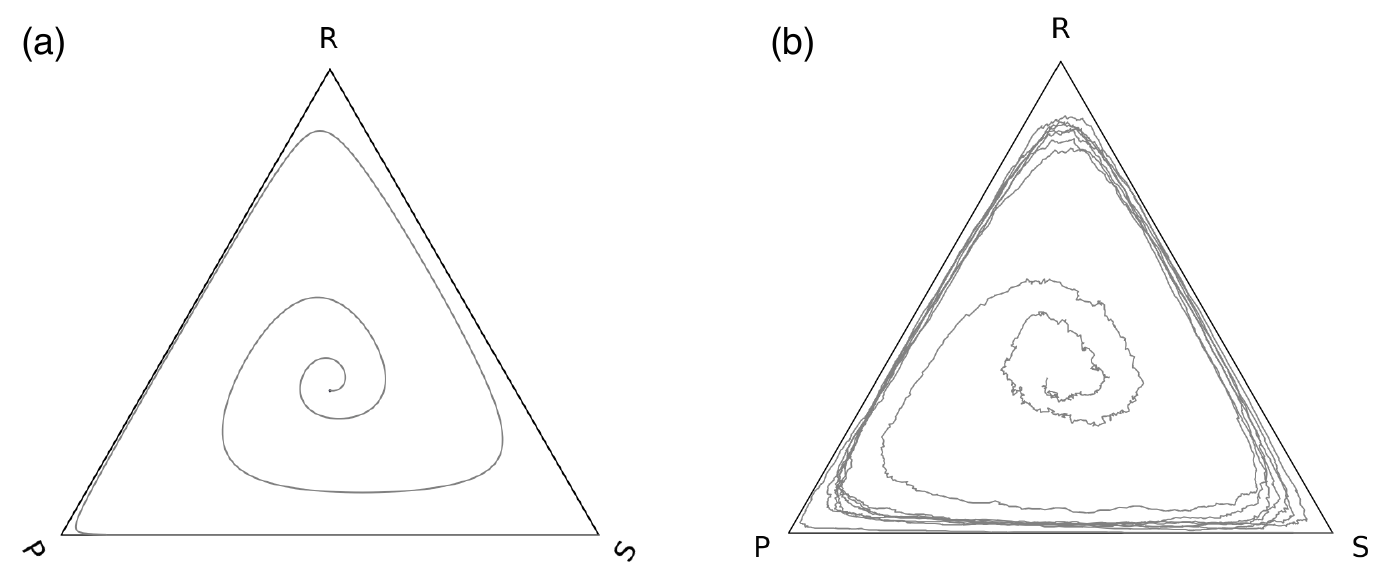} 
   \caption{Replicator equation modeling  for the Rock-Paper-Scissors game with a repulsive fixed point at population fractions $(x_R,x_P,x_S)=(1/3,1/3,1/3)$. (a): infinite population size (replicator equation), with initial condition at the fixed point. (b): agent-based modeling (finite population size $N=1,024$), Moran process (see Box 3), population frequencies plotted every 25 generations. In this case, the population ended up on the fixed point $x_S=1$.}
   \label{fig:RPS}
\end{figure}

A typical example illustrating these ideas among three-strategy games is the Rock-Paper-Scissors (RPS) game, one of the 38 possible three-strategy games described by Zeeman~\citep{Zeeman1980}. A typical payoff matrix for this game (in normal form with vanishing diagonal, and with internal fixed point $(1/3,1/3,1/3)$, that is, equal population fraction for all strategies) is given by
\be
\bordermatrix{\mbox{} & {\rm R} & {\rm P} & {\rm S}  \cr
                           {\rm  R}      & 0  & -2 & 1 \cr
                            {\rm P}      &1 &  0 & -2 \cr
                            {\rm S}      & -2 & 1 &0 }\;.  \label{rps-payoff}
\ee   
This fixed point, however, is unstable: the trajectories push outward from the central point, and soon enough the population (in the infinite population approximation) moves from all scissors, to all rock, to all paper (a trajectory known as a heteroclinic orbit because it connects the unstable pure strategy fixed points, see Fig.~\ref{fig:RPS}a). 

Such orbits are impossible for a finite population, however. In these, once a type is extinct it will not return (see Fig.~\ref{fig:RPS}b). In finite populations playing the same game (defined by the payoffs~(\ref{rps-payoff})) only a single strategy will survive, but which one ultimately remains is random. 
Note that extinction can happen due to the stochasticity introduced by the finite population even for the Rock-Paper-Scissors game with an {\em attractive} interior fixed point (obtained from Eq.~(\ref{rps-payoff}) by, for example, flipping the signs in that matrix)~\citep{ClaussenTraulsen2008,Adamietal2012}.

We do not mean to imply here that no results can be obtained analytically for finite populations--quite to the contrary (see, e.g.,~\citep{TraulsenHauert2009} for a comprehensive review). For example, it is possible to calculate the probability that a particular strategy can invade an existing (clonal) population, as a function of the size of the invading ``clan", for arbitrary $2\times2$ games, that is, games between two strategies defined by payoff matrices of the type (\ref{game}). In certain limits (for example, very small mutation rates and a pre-defined set of strategies as discussed below), the finite-population dynamics can be described by a Markov process that describes the transition probability between each strategy. The population trajectories can then be obtained by simply iterating the Markov matrix, and the fixed point can be obtained by calculating the (left) eigenvector of the Markov matrix (see, e.g.,~\citep{NowakSigmund1993,TraulsenHauert2009}). Thus, finite population sizes by themselves are not a reason to abandon mathematics in favor of agent-based simulations. But as we will see, when finite populations are coupled with a number of other realistic aspects of evolving populations (in particular higher mutation rates), agent-based simulations are essential in order to understand the evolutionary fate of populations that play games. 

\subsection*{Mutations} \label{sec:muts}
If mutations constantly produce new strategies, analytical methods must fail because they cannot keep track of the persistent production of novelty. In the previous section, we studied finite populations using agent-based modeling, but strategies were not mutating. The game dynamics simply determines which strategy (or set of strategies) should survive, changing the frequencies $x_i$ accordingly. But in Darwinian evolution strategies can change via random mutations, and we can implement this process in our agent-based simulations. If mutations are possible, the replicator equations (for the case of infinite populations) have to be replaced by the {\em replicator-mutator} equations: this is Eigen and Schuster's quasispecies model~\citep{Eigen1971,EigenSchuster1979}. Unfortunately, the quasispecies equations are exactly solvable only for very specific fitness landscapes and mutation processes. However, in the limit of small mutation rates and a finite population size the evolutionary dynamics can (as we mentioned above) be described by a Markov process whose stationary distribution can be calculated exactly~\citep{TraulsenHauert2009}.   

In case there are only a few possible strategies, the effect of mutations mainly modifies the finite-population effect, where (in the absence of mutation) strategies can go permanently extinct. Mutations can resurrect strategies that went to extinction, which could be advantageous for the population if payoffs (dictated by the environment, for example) change. In this case (just a few possible strategies), the effect of mutations can be simulated by simply introducing a constant (but small) rate of strategy influx that ensures each of the strategies is re-introduced into the population (see, e.g.,~\citep{NowakSigmund1993}).

In the limit of a large (or even infinite) set of possible strategies, it is not possible to create a constant flux of all strategies into the population, neither would that be desirable. In a realistic finite population at a finite mutation rate, only a tiny subset of all possible strategies ever coexist, and only those strategies that are mutations of the existing set can enter the population. Because the existing set ever changes (due to extinctions and the emergence of new strategies), the set of mutant types also constantly changes. In such a setting, strategies that become extinct are unlikely to be resurrected, and as a consequence the population is dynamic and continues to explore strategy space. Such an evolutionary dynamic can only be implemented by providing a genetic basis for each strategy, so that the mutations of the genes create progeny whose strategy will be similar to that of their ancestors. This is the basis of agent-based simulation in evolutionary game theory. 

{\bf In realistic biological populations, multiple beneficial mutations exist at the same time.} 
In evolutionary theory, it is customary to distinguish two different adaptive regimes, dictated to a large extent by the rate of mutation. If the mutation rate is so low that it is unlikely that more than two 
different types (the resident type and a candidate mutation) ever coexist in a population (this happens when the {\em mutation supply rate}, given by the product of population size and mutation rate is significantly smaller than 1) then the evolutionary history of a population can be described by a sequence of substitutions that each went to fixation individually in a homogeneous background. As discussed before, the probability of any of the two strategies to go to fixation can then  be described analytically (see, e.g.,~\citep{TraulsenHauert2009}).  This regime is often called the ``strong selection-weak mutation" (SSWM) limit (see~\citep{SniegowskiGerrish2010} for a review and a discussion of the empirical evidence for rates of mutation).  In this regime, the likelihood that a particular variant wins the evolutionary race and goes to fixation can be calculated mathematically, even at finite population size. 

The other extreme of evolutionary dynamics is the ``weak selection-strong mutation" (WSSM) limit, where multiple variants with beneficial mutations are coexisting in the population and where the standard mathematical theory of fixation of single beneficial mutants~\citep{Ewens2004} does not apply. A rule of thumb that is often used to separate these two regimes is simply $N\mu\sim 1$, where $N$ is the effective population size, and $\mu$ is the rate of beneficial mutations, implying that per generation only a single new beneficial mutation arises, which then has a chance to become dominant without any competition from other mutants. However, a more accurate estimate of the line that separates the two regimes suggests that for SSWM to hold sway, $2N\mu\ll(\ln Ns/2)^{-1}$, where $s$ is the average beneficial effect of a mutation~\citep{SniegowskiGerrish2010} (see also~\citep{DesaiFisher2007}). For realistic benefits $s$, this dividing line is approximately $\mu\approx 1/N^2$, which we confirm empirically below.   

For bacterial populations, the per-site mutation rate is of the order $10^{-10}$ per nucleotide~\citep{Barricketal2009}, translating to a per-genome mutation rate (for {\it E. coli} bacteria) of about $5\times 10^{-4}$, that is, only 5 out of 10,000 bacteria produced carry at least one mutation, and roughly only one out of 50 of those mutations are beneficial~\citep{SniegowskiGerrish2010}. Given the typical size of bacterial populations of the order of $10^6-10^8$, a beneficial mutation rate of $10^{-5}$ implies that many beneficial mutations will coexist at any one time in a population, even though only a few of them will make it on to the line of descent as they inevitably will interfere with each other~\citep{GerrishLenski1998}. Using $N=10^7$ and a mean beneficial effect of 1\%, $(\ln(Ns/2))^{-1}\approx 10^{-1}$, while $2N\mu\approx200$ (using a beneficial mutation rate per genome per generation of about $10^{-5}$~~\citep{SniegowskiGerrish2010}). Thus, even bacteria with the arguably smallest mutation rate are solidly in WSSM territory, where  multiple mutations (sometimes multiple mutations on the same individual~\citep{Maddamsettietal2015}) fight it out. 

Where does this lead your standard EGT simulation? To maximize evolutionary speed, most simulations operate at a genomic mutation rate of about 1. In early adaptation, most of these mutations are beneficial, and even if we assume that only 10\% are, then $2N\mu\approx 200$ (for a typical population size $N=1,000$). Given that the average beneficial mutation in such simulations has an effect of (roughly) about 10\%, $(\ln(Ns/2))^{-1}\approx 0.25$ in EGT simulations, putting them also decidedly in the WSSM regime. As it is well known that fixation theory does not work in this regime (although some analytical results can be had for other population observables~\citep{Desaietal2007}), we should expect that EGT simulations in this regime give results differing from mathematical approaches, and we will encounter such examples in the section on mutational robustness below. 

\subsection*{Stochastic strategies}
If we think about real agents making decisions in an uncertain world (be it microbes or day traders), it is the rarest individual who makes decisions deterministically. More often than not, an agent is described by a {\em probability} to make a decision. 
In games with two strategies, Maynard Smith showed that for a game with an attractive interior fixed point (a snowdrift game with $a,b>0$) a {\em probabilistic} strategy with probabilities given by the equilibrium population fractions of the deterministic game is in fact evolutionarily stable, that is, an ESS of the game~\citep{MaynardSmith1982}. However, this general result does not translate to games with more than two strategies. While the fixed point for stochastic strategies will always coincide with the mixed strategy ESS (the stable population fraction of pure strategies), there is no general theory that predicts the stability of said fixed point~\citep{Hines1980,Zeeman1981,Thomas1985,Bomze1990,Cressman1990}, mostly because the stability criteria involve only strategies close to the fixed point.
We can see this readily by analyzing a three-player game that is general enough to include the standard Rock-Paper-Scissors game (both with an attractive and a repulsive fixed point), and a number of other three-player games classified by Zeeman~\citep{Zeeman1980}.

The ``Suicide Bomber" (SB) game is modeled after the dynamics of bacterial populations in which a small fraction of bacteria commit suicide by triggering an explosion that sprays a bacterial toxin into the surrounding area. While the exploding bacterium is killed, its kin (who carry a resistance gene to the toxin) profit from the suicide, because it removes non-kin bacteria from the food source as those do not carry the resistance gene~\citep{LenskiVelicer2000,ChaoLevin1981,Kerretal2002}. But the dynamics are complicated: the wild-type strain ``00" that carries neither the toxin gene T nor the resistance gene R will be outcompeted by the suicide bomber ``RT"  strain, because of the advantage that explosion confers on the bomber's kin. However, RT carries a double disadvantage compared to the wild-type that does not carry either gene, because toxin production and resistance are both costly. The RT strain, as a consequence, can be invaded by a strain that has lost the toxin production gene (a strain ``R0"), because it is a cheater that does not suffer from toxin exposure, yet does not pay the cost of carrying the toxin. Once R0 dominates the population, it can be invaded by a wild-type 00, because in the absence of toxin production, carrying the resistance gene is a useless luxury. Of course, once 00 dominates, it is again vulnerable to invasion by RT, and the cycle resumes in what seems like a never-ending game of Rock-Paper-Scissors.

{\bf Stochastic strategies can be stable even if the corresponding mixed state is unstable}. A quantitative analysis reveals a subtle dependence of the game dynamics on the relative size of the costs and benefits of the R and T gene~\citep{Adamietal2012}, revealing that the game dynamics can belong to one of seven of Zeeman's 38 possible three-strategy games. In particular, the fixed point of the RPS game can be either attractive or repulsive. In Fig.~\ref{fig:stab} below, we show the phase portrait of the SB game for the repulsive RPS game ($\omega<\veps, \veps<1$, see the payoff matrix for this game in Box 2), using deterministic strategies and an infinite population (solved using the replicator equation) on the left, and using agent-based simulations of stochastic strategies on the right. On the left diagram, each point on the trajectory represents a set of population fractions (the mixed state), while for the agent-based simulation on the right, the trajectory represents an (average) line of descent of the probabilities to engage in the three different plays. We observe that the trajectory on the left spirals outwards, away from the repulsive fixed point indicated by the yellow arrow, while on the right, the trajectory moves {\em towards} the fixed point (even though it does not appear to quite reach it). Thus, a repulsive  fixed point for deterministic mixed strategies has turned into an attractive fixed point for stochastic strategies, while the location of the fixed point appears to be unchanged.
\begin{figure}[htbp] 
   \centering
   \includegraphics[width=4in]{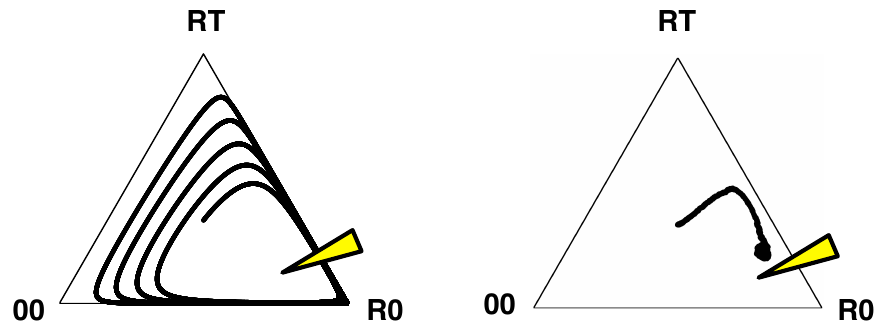} 
   \caption{Population fraction trajectory for an SB game with $\omega=0.125$ and $\veps=0.75$. Population fractions of a deterministic game modeled with replicator equation (left), and average probabilities on the line of descent on the right. Simulations are started with equal fraction of three types (left) or with a population of 1,024 strategies with probabilities (1/3,1/3,1/3) on the right.}
   \label{fig:stab}
\end{figure}
\vskip 0.25cm
\noindent\shadowbox{
\begin{minipage}{6.3in}
\centerline{\bf Box 2: Suicide Bomber game}
\mbox{}\vskip 0cm
\noindent In this game, three strategies 00 (the ``wild-type"), RT (the ``bomber", who carries the toxin gene T as well as resistance to it R), and R0 (the ``cheater", who carries resistance but no toxin) battle against each other, with dynamics dictated by the relative value of the benefit $\veps$ and the cost $\omega$ (we assume here an equal cost for both the toxin and the resistance gene). A fourth possible strategy (0T) is not viable, because carrying the toxin gene without resistance to it is generally a bad strategy.  In practice, only a few percent of the toxin-carrying organisms explode~\citep{ChaoLevin1981}, reducing the cost to the carrier.
The payoff matrix is given by 
\be
\bordermatrix{\mbox{} & {\rm 00} & {\rm R0} & {\rm RT}  \cr
                           {\rm  00}&   1   & 1  & 0 \cr
                            {\rm R0}&1-\omega&1-\omega &  1-\omega \cr
                            {\rm RT}&1-2\omega+\veps & 1-2\omega & 1-2\omega  }\;. \label{3play}
\ee      
The seven possible games that exist for this payoff matrix can be described by {\em phase portraits} that sketch the expected population trajectories. 

 \centering
   \includegraphics[width=3in]{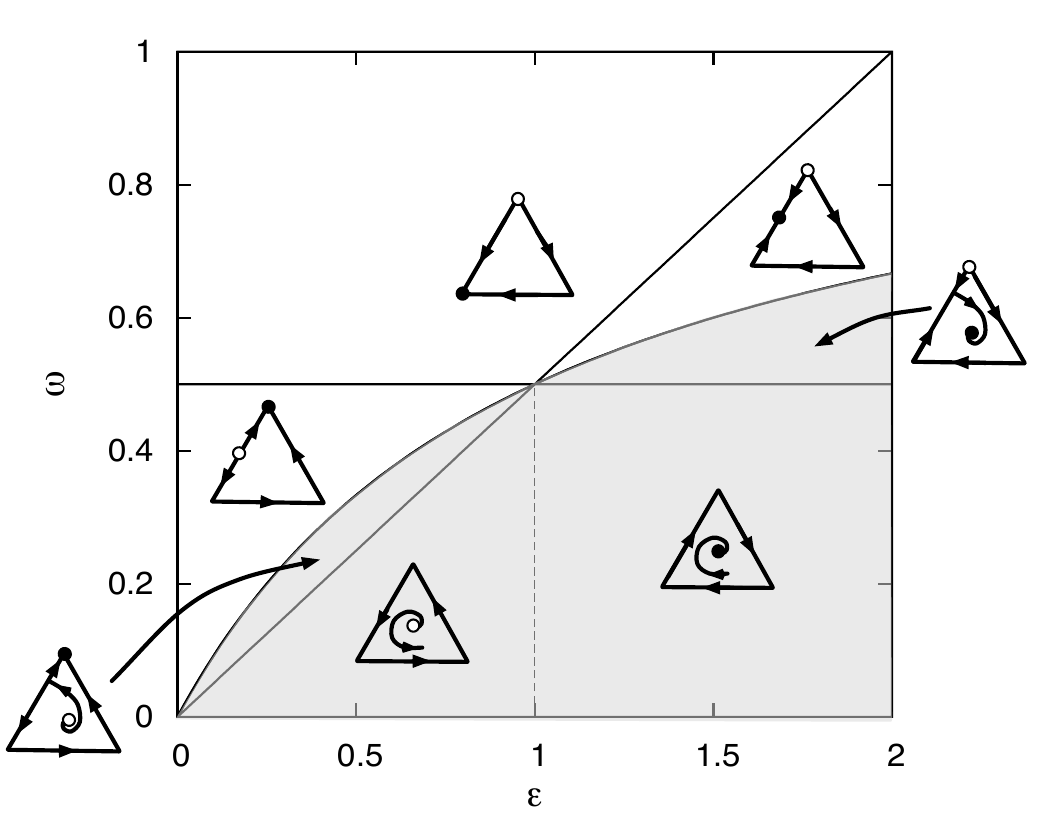} 
   \captionof{figure}{Phase portraits of stable dynamics for the SB game using payoffs (\ref{3play}). The shaded parameter region $\omega<{\veps}/({\veps+1})$ has an interior fixed point that can be repulsive or attractive. In Zeeman's phase portrait pictograms~\citep{Zeeman1980}, arrows denote the flow on the boundary of the simplex, solid circles are attractors and open circles are repellers. All fixed points on the boundary and the interior are indicated. Modified from~\citep{Adamietal2012}.}
  \label{fig:Zeeman}

\end{minipage}
}
\subsection*{Evolutionarily stable sets}
Consider for a moment a two-player game described by the payoff matrix (\ref{game}) in the ``snowdrift" regime, where both $a,b>0$.  In that case, none of the ``pure" strategies C or D are an ESS, but the {\em mixed strategy} M (a mixture of the two strategies C and D) is, with frequencies $\frac{a}{a+b}$ and $\frac{b}{a+b}$ respectively. 

\begin{figure}[htbp] 
   \centering
   \includegraphics[width=3in]{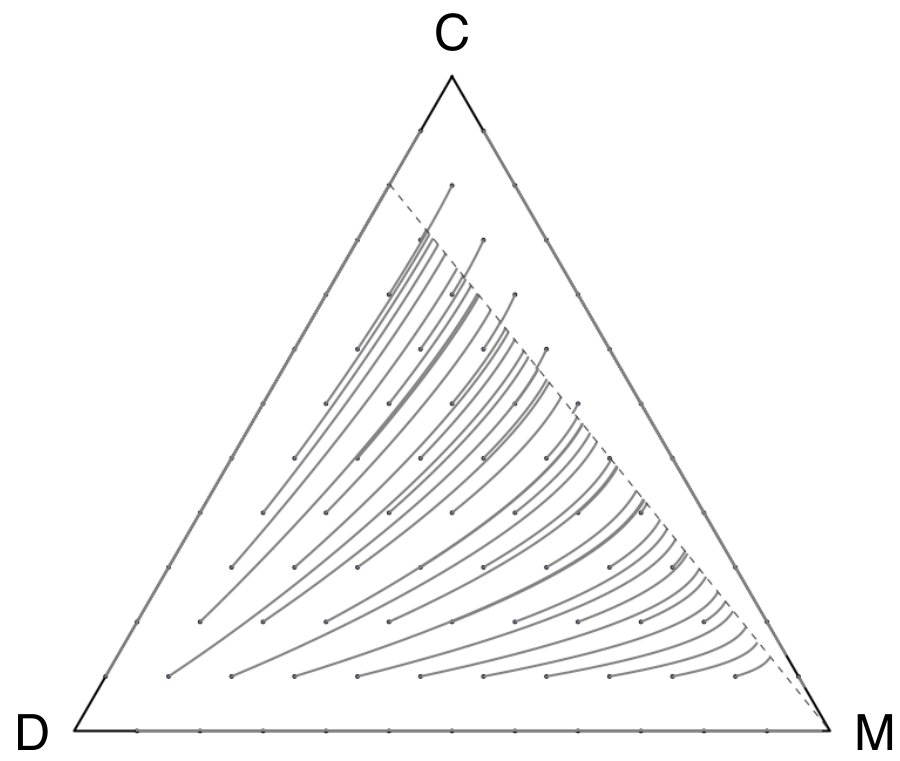} 
   \caption{Trajectories of the population fraction of the game defined by Eq.~(\ref{ess-payoff}) with $a=1$ and $b=0.2$. The dashed line represents the ES set, while the solid lines show the population trajectories that begin at the solid dots (different initial conditions), and move towards the ES set.}
   \label{es-set}
\end{figure}

The mean payoff of strategy M against pure strategies C and D can be calculated, so we may ask: ``What happens if we play the pure strategies C and D against another strategy that has the same payoffs as the strategy mixture?" For example, the strategy M earns $b^2/(a+b)$ against C, and $a^2/(a+b)$ against D. What is the dynamics when a pure strategy with these exact payoffs is thrown ``into the mix"? We can investigate this by studying the payoff matrix 
\be
\bordermatrix{\mbox{} & {\rm C} & {\rm D} & {\rm M}  \cr
                           {\rm  C}      & 0  & a & ab/(a+b)  \cr
                            {\rm D}      &b &  0 & ab/(a+b)  \cr
                            {\rm M}      & b^2/(a+b) & a^2/(a+b) & ab/(a+b)}\;,  \label{ess-payoff}
\ee  
which is easily deduced from the payoffs of M against C or D. Note that the M column could as well be zero, because subtracting the same constant from every element in a column does not change the game dynamics.
It turns out that in such a game isolated fixed points turn into {\em stable sets}: M can be stable in the background of C and D at arbitrary frequencies, but given a particular frequency of M (say, $r$), the frequencies of C and D are fixed at $a(1-r)/(a+b)$ and $b(1-r)/(a+b)$ respectively. Such {\em Evolutionarily Stable Sets} (ES sets) were first discussed by Thomas~\citep{Thomas1985a,Thomas1985b} and are studied in detail by Weibull~\citep{Weibull1995}. A deterministic (pure) strategy M with payoffs as defined in (\ref{ess-payoff}) will form ES sets, because it is neutral with respect to the other strategies (see Fig.~\ref{es-set}). Along the neutral line in Fig.~\ref{es-set}, no one set of strategies is better than another, forming a {\em ridge of attraction} in the phase portrait (see also~\citep[p. 75]{HofbauerSigmund1998} and~\citep{Cressman2003}). 

This analysis of ES sets did not require agent-based methods. But we could now go further and ask, what is the dynamics of a {\em probabilistic strategy} that can play either of the three strategies C, D, or M with probabilities $p$, $q$, and $r$? As discussed above, we expect the fixed point of the deterministic theory to predict the stable point of the probabilistic theory, which would imply that the entire dashed line in Fig.~\ref{es-set} should also be fixed points of the {\em probabilistic dynamics}. The deterministic theory does not predict the stability of the probabilistic ES set, but previous experience~\citep{Adamietal2012} suggests that the set should be attractive. 
 
{\bf The encoding of decisions into genetic loci can affect evolutionary trajectories.}  
When testing these predictions with agent-based methods, several decisions in designing the simulation can have a significant impact on the results. For example, because the probabilities $p$, $q$, and $r$ are changed via a discrete mutational process, the nature of that process will affect the population dynamics. If the probabilities are implemented as continuous variables (rather than discretized to a particular resolution), we could mutate either by replacing the given probability by a uniform random number (``global" mutations), or we could change the probabilities either up or down by a uniform random number from a distribution spanning a particular percentage (``local" mutations). In the latter case, care must be taken so as to remain within the boundaries of a probability. At the same time, it is not possible to update all three probabilities independently, as they must sum to one. Thus, if we implement the underlying genetics of the process in terms of three loci (one for $p$, one for $q$, and one for $r$), then mutating one locus will necessarily affect both other loci (we refer to this implementation as the ``3-gene" implementation). If we instead implement the genetics in terms of two independent loci (say, $p$ and $q$, the ``two-gene" implementation), then the value of the third probability is determined automatically. All these design decisions affect the population dynamics, as we will now see. 

\begin{figure}[htbp] 
   \centering
   \includegraphics[width=6.5in]{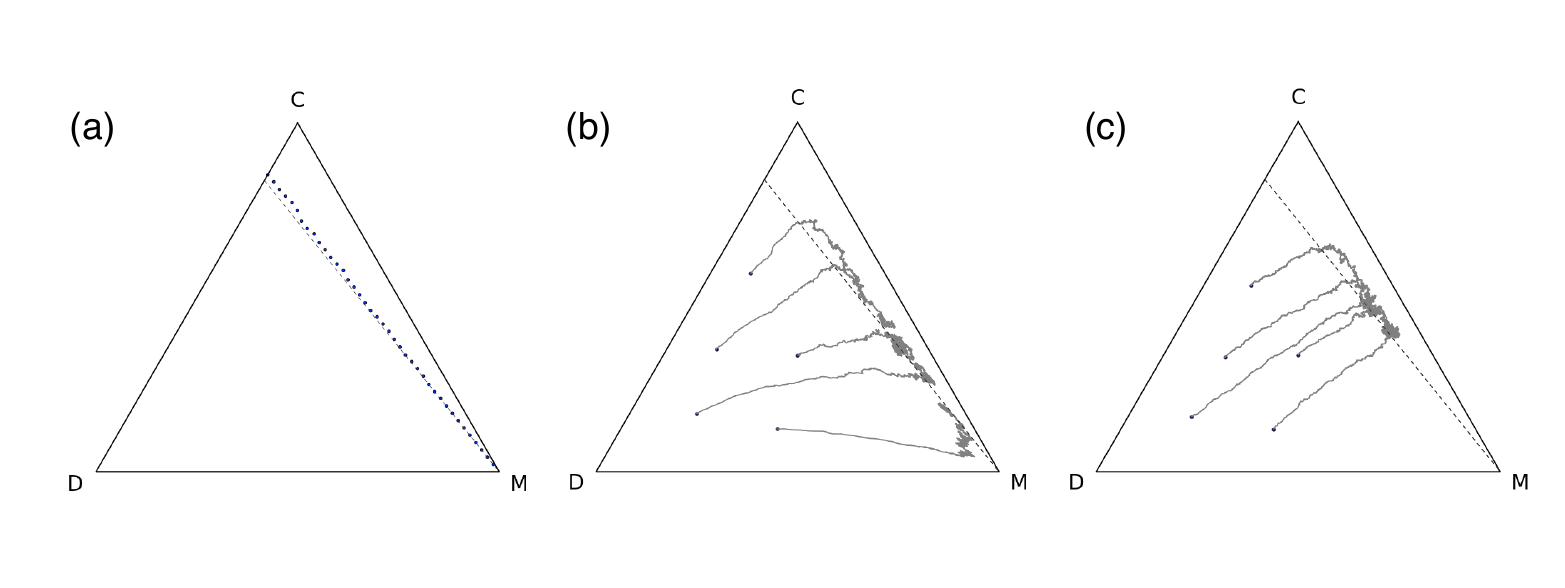} 
   \caption{ (a): Averaged end points of 1,000 trajectories (averaged $p$ and $q$, for slices with fixed $r$). End points are defined to be the most recent common ancestor of the population on the line of descent, usually this point is around generation 900 (of 1,000). Averaging the end points without fixing $r$ leads to a single point, as the law of large numbers implies that the mean of the ES-set in the $r$-direction equals 0.5 (not shown). The payoff matrix  in (\ref{ess-payoff}) has $a=1,b=0.2$. (b): Average trajectories of lines of descent for probabilistic strategies with different initial conditions, using the ``3-genes" encoding (see text). The dashed line represents the predicted ES-set for deterministic strategies shown in Fig.~\ref{es-set}. Average of 50 trajectories per line of descent, obtained from populations with 1,024 agents evolving at a mutation rate of 10\% per locus (translating to 20\% per chromosome) for 3,000 generations. (c): Average trajectories of lines of descent for probabilistic strategies using an encoding where each of the two probabilities $p$ and $q$ is encoded independently, thus fixing $r$ (the ``2-genes" encoding). Other parameters as in (b).} 
   \label{es-set-prob}
\end{figure}

In Fig.~\ref{es-set-prob}a, we show the average {\em end points} of the set of probabilities $(p,q,r)$ on the evolutionary ``line of descent" (see Box 3). Note that only the average trajectory (averaged over 1,000 trials) starting at the same initial state $(p(0),q(0),r(0))$ is smooth: each single trajectory itself is jagged. In this run, when probabilities are mutated they are changed at most $\pm 5\%$ from their current value, which can give rise to significant jumps within the triangle. We see the trajectories to reach the ES-set in Fig.~\ref{es-set-prob}b for the ``3-genes" encoding, where mutations in one of the loci affects the other two probabilities (as only two of the three probabilities are independent). 
The trajectories differ when the decisions are encoded in two independent loci (Fig.~\ref{es-set-prob}c), while the end-points are of course the same.

We thus see that agent-based methods can also be used in game-theoretic problems with an ES-set. The trajectory of probabilities on the line of descent follows the population fractions of the pure strategy only approximately, something that we had also noticed for trajectories in a game with an isolated fixed point in Fig.~\ref{fig:stab}. We ought not to worry about this, however, as we recall that the theory of evolutionary stability of stochastic strategies~\citep{Hines1980,Zeeman1981,Thomas1985,Bomze1990} only predicts the location of the fixed point, not its stability or the trajectory that game dynamics uses to reach the fixed point. But it is clear that the population dynamics in EGT must depend both on genotype-phenotype mappings as well as mutational mechanics, elements that are difficult to study using analytical methods.

\vskip 0.25cm
\noindent\shadowbox{
\begin{minipage}{6.3in}
\centerline{\bf Box 3: Methods in Agent-based Simulations}
\mbox{}\vskip 0cm
\noindent 
In agent-based simulations, strategies are encoded within genes (sometimes called loci) that make up the genotype or chromosome, and these genes are evolved using a Genetic Algorithm~\citep{Michalewicz1996}. While GAs traditionally utilize recombination between genotypes as a method of randomization, this is rarely used in EGT simulations as the number of loci is usually low. In the stochastic game-theoretic application discussed here, the genes are just sets of probabilities, one for each action that an agent can take. In an unconditional $2\times2$ game (two-player, two strategy game), a single gene encoding the probability to play C, for example, suffices. In a conditional $2\times2$ game, 5 genes are needed (four encoding the conditional moves and one for the first move of the player, which is unconditional). In a memory-2 game (where players can make their decision based on the previous two moves) we need to add another 16 loci, for a chromosome with 21 genes~\citep{Iliopoulosetal2010}). For the unconditional Public Goods (PG) game (i.e., the game where strategies' decisions do not depend on previous play), only a single gene is needed unless punishment is used, which adds another locus. The conditional PG game where each agent's decision depends on the focal player's own play and all $k$ players in his group would require $2^{k+1}$ genes, but most approaches instead make an agent's decision depend only on its own prior play and the number of cooperators in its group, requiring only $2k$ genes.

Every one of the agents in the population plays a fixed number of games before the population is updated, and the payoff accumulates for each. In a well-mixed population, agents play opponents that are chosen at random from the population, while in a spatial simulation an agent plays its nearest neighbors on a grid. Typically, a fixed {\em random} fraction $R$ of the population is then removed. This fraction can range from removing only a single player ($R=1/N$, where $N$ is the population size) in a procedure called the {\em death-birth Moran process}~\citep{Moran1962}, to removing the entire population ($R=1$)--this is called the Wright-Fisher process for asexual haploids~\citep{DonnellyWeber1985}. The fraction of the population that is removed each update determines the number of games each player engages in during their lifetime, and therefore plays an important role in the population dynamics. As $R$ is the fraction removed each update, the inverse $1/R$ (given the population size) determines the average number of updates each individual survives, and since the number of games per update is constant, $1/R$ determines the average number of games each individual plays before they are removed. 

Sometimes, a ``strategy imitation" mechanism is used for selection, where strategy $i$ receiving payoff $E_i$ is replaced with probability $P_{i\to j}=(1+{\rm exp}(E_i-E_j)/K)^{-1}$, where $E_j$ is the payoff of any of the opponents, and $K$ is a measure of noise (see, e.g.,~\citep{SzaboFath2007}). Even though the fixation probability of strategies in such a ``Fermi process" (after the Fermi function $(1+e^{x/T})^{-1}$) differs from that of the Moran or Wright-Fisher process  (see, e.g.,~\citep{TraulsenHauert2009} for a lucid comparison of the three processes), the ultimate evolutionary outcomes are the same.

If random players are removed, selection must occur on the birth process, and indeed the number of offspring each strategy receives is determined by its ranking by the relative fitness. After the population is filled-up back to its original size after culling, mutations are applied to the chromosomes based on the prevailing mutation rate. Because mutations are Poisson-random, the probability that a genome suffers $n$ mutations is given by the Poisson distribution $P(n)=\mu^ne^{-\mu}/n!$ where $\mu$ is the mutation rate (mean number of mutations per chromosome per update). Note that if the mutation rate per locus is $\lambda$, the per-chromosome mutation rate is $\mu=\lambda L$, where $L$ is the number of loci or genes.

In order to track the evolution of the chromosome, we can reconstruct the ancestral {\em line of descent} (LOD) of a population by picking a dominating type (for example, the one with the highest fitness at the end of the simulation), and identifying its ancestor, then the ancestor's ancestor and so forth, arriving finally to the type used to seed the simulation. As there is no sexual recombination between strategies, each population has a single LOD after moving past the most recent common ancestor (MRCA) of the population. The LOD recapitulates the evolutionary history of that particular simulation, as it contains the sequence of mutations that gave rise to the successful strategy at the end of the run (see, for example~\citep{Lenskietal2003,Ostmanetal2012}).
\end{minipage}
}

It is interesting to note that different genetic encoding strategies result in different mutation-induced trajectories through the phase-portrait. This is because although both 2-gene and 3-gene encoding strategies have the same evolutionary pressure (selection gradient), they differ in the amount of gene-to-gene interaction (epistasis) as well as total effect on the final 3 probabilities. That is, a change in one of two interacting genes (b) alters the resulting three traits differently than a change in one of three loosely interacting genes producing the same traits (c). However, all trajectories ultimately end up on the ES-set, and then drift along that line. 
However, we should keep in mind that in evolutionary biology, the evolutionary trajectory is at least as interesting as the endpoint(s), so differences in encoding that affect the trajectory are worth studying.
\subsection*{Conditional strategies}
In the previous sections, we considered strategies that made decisions unilaterally. In more realistic situations, agents prefer to make {\em informed} decisions, meaning that they will utilize external clues to modulate their decisions. We will now study {\it communicating} strategies in the Prisoner's Dilemma: the perennial ``go-to" game to study the emergence and maintenance of cooperation in EGT (one of the four Zeeman classes of the two-strategy games, see Box 1). The name ``Prisoner's Dilemma" stems from a hypothetical situation where two thieves who have committed a crime flee from the scene, and are intercepted by the police who are convinced that one of them is the perpetrator, but are unable to pin the crime on any one of the two. They haul in the thieves, and interrogate them {\em separately}. The idea here is that should one rat out the other, the police have their culprit and the other goes free. A prisoner who stays silent is said to be ``cooperating" (with the other prisoner), but they each have an incentive to squeal. The standard payoff matrix for this game is 
\be
\bordermatrix{\mbox{} & {\rm C} & {\rm D}  \cr
                           {\rm  C}      & R  & S \cr
                            {\rm D}      &T &  P   }\;,  \label{PD-payoff}
\ee   
and for the game to be in the ``PD" class, we must have $T>R>P>S$.  Several different payoff matrices are commonly used, the most widespread among them is perhaps Axelrod's $(R,S,T,P=3,0,5,1)$. Another payoff matrix in this class that is often used is  $(R,S,T,P=2,-1,3,0)$, sometimes called the ``donation game"~\citep{Hilbeetal2013}. When re-scaled by adding one to the C column and subtracting one in the D column, we can see that the donation game is just the Axelrod payoffs with a somewhat weaker temptation payoff $T=4$. 

In PD, the rational strategy (the Nash equilibrium) is to defect, giving each of the players a payoff of 1, even though they obviously would have done better if they had both cooperated so as to reap a payoff of 3 each: thus the dilemma. For over thirty years this dilemma has been upheld as epitomizing the ``paradox of cooperation", ostensibly because it is hard to imagine how cooperation could evolve via Darwinian principles that favor short-term gains over long-term potential gains, principles that therefore should favor defection rather than cooperation. However, this conclusion only holds in the absence of communication, and anyone would be hard-pressed to find a single example of cooperation in nature that does not involve some form of communication\footnote{We do not count here mutualistic associations in which there is no dilemma. It is possible to evolve cooperation purely via spatial reciprocity, but it is not clear whether assortment~\citep{FletcherDoebeli2009} by space alone (without assortment via communication) can ever be stable over the long-term.}. And if the prisoners had not been interrogated separately, the would surely form a pact ensuring that both get the lesser sentence (the higher payoff 3). 

It is easy to extend the standard game to allow for communication. Strategies that take information into account are called ``conditional strategies", because the agent's move is conditional on symbols that they obtain. A good source of information is past play (the play of the opponent, but also the agent's own past moves) so as to be able to gauge the opponent's play in relation to one's own. It is clear that for such conditional strategies to be effective, the game has to be repeated, that is, iterated.

{\bf Communication is essential for cooperation.} The quintessential conditional strategy in the iterated PD (IPD) is (because of the sheer amount of literature devoted to it)  a strategy called ``Tit-for-Tat" (TfT). This strategy, submitted by Anatol Rapoport to Axelrod's first tournament that pitted different computer strategies against each other~\citep{Axelrod1984}, rose to fame because it was able to amass the highest score in the tournament\footnote{Whether TfT should have been declared the winner was recently challenged~\citep{Rapoportetal2015} because TfT profited from a peculiar tournament structure, winning the tournament without winning a single pairing.}.
TfT cooperates if the opponent cooperated in the previous move, and defects if that is what the opponent just did. We can describe a strategy in terms of conditional probabilities that take the past move of the opponent as well as the past move of self into account as\footnote{The first symbol after the bar (read: ``given") refers to the agent's own last move, while the second symbol refers to the last move of the opponent. So, for example, $p({\rm C|DC})$ is the probability for the agent to cooperate given that the agent just defected and the opponent just cooperated. The probability to defect given these moves is just $p({\rm D|DC})=1-p({\rm C|DC})$.}
\be
\vec P= \left(p({\rm C|CC}),p({\rm C|CD}),p({\rm C|DC}),p({\rm C|DD})\right)\equiv(p_1,p_2,p_3,p_4)\;. \label{stoch}
\ee
In terms of these probabilities, the strategy TfT is $\vec P_{\rm TfT}=(1,0,1,0)$. Strategies that take only the last move into account are called ``memory-one" strategies. It is straightforward to extend the concept of conditional strategies to those using longer memories. 

Of course, TfT is not a probabilistic strategy, nor do its actions depend on its own past moves, but we introduced this notation so as to be able to discuss the more general stochastic ``memory-one" strategies, introduced by Nowak and Sigmund~\citep{Nowak1990,NowakSigmund1990}. The dynamics of this infinite set of strategies cannot be described in its entirety purely mathematically, simply because the mathematical calculation of optimal strategies is just too cumbersome (for an example, see Eq.~(15) in~\citep{AdamiHintze2013}, also see~\citep{Akin2012} for a classification of all ``good" strategies). 

If mathematical methods will not tell us which stochastic memory-one strategy is favored by evolution, agent-based methods are there to tell the story. If the question is: ``What is the optimal stochastic conditional strategy in PD that is favored by evolution?", the answer most definitely is: ``It depends", as has already been argued long ago~\citep{BoydLorberbaum1987}. By using an approach where the four probabilities (\ref{stoch}) (supplemented by a probability that decides the first unconditional move in an encounter) are encoded genetically, Iliopoulos et al.~\citep{Iliopoulosetal2010} showed that what strategy ultimately dominates the population depends on the environment. For example, high mutation rates favor strategies that defect, because the high rate of mutation tends to modify the opponent (often while engaged in play), rendering the opponent less reliable. If the interaction between two players is viewed as a communication channel (with maximal capacity of two bits: one bit for the channel between past and future play for the agent and one bit for the channel from the opponent~\citep{Mirmomeni2015}) then mutations act like noise in the channel. Increasing channel noise leads to a decrease in the channel capacity, and thus to a decrease in the amount of information exchanged per interaction.  Similarly, if the population turns over quickly because a large fraction of players is replaced at each generation, any one player cannot rely on playing the same opponent, and opts instead to defect. Because the population size often determines how many games any strategy plays with its opponent before it or the opponent is removed, the population size itself will also determine the winning strategy. What we witness then is a {\em phase transition}~\citep{Iliopoulosetal2010}  between cooperative and defective behavior that is driven by the mutation rate or the mean length of an iterated game. High mutation rate and high replacement rate (small population size) create high uncertainty (less information) about the opponent, and results in a cautious population that favors defection. 

Significant mathematical progress in elucidating the structure of the space of stochastic memory-one strategies was made recently, after Press and Dyson discovered that a subset of these strategies is able to take advantage of their opponents~\citep{PressDyson2012}. Even though strategies of this type had been mentioned in the literature before~\citep{Boerlijstetal1997}, this discovery emphasized that the set of stochastic memory-one strategies is far from being fully understood. Press and Dyson discovered a set of conditional strategies that communicated with their opponents in a nefarious manner: they used the opponent's choices in such a manner as to {\em manipulate} the opponent. This class of strategies, termed ``Zero Determinant" (ZD) strategies after a nifty mathematical trick discovered by Dyson, in reality contains two classes: the selfish types that take advantage of their opponent, and compliant ones (also termed ``generous strategies"). Among the selfish strategies there are two main types: the ``Equalizers" that manage to force a fixed payoff onto the opponent, and the ``Extortionists" that fix the relative payoff between player and opponent in such a manner that if the opponent changes its strategy so as to get a better payoff, the extortioner always receives commensurately more.  The compliant ZD strategies are in fact not nefarious (these were not studied by Press and Dyson) and were discussed in detail first by Stewart and Plotkin~\citep{StewartPlotkin2012,StewartPlotkin2013}, and also by Akin~\citep{Akin2012} and Hilbe et al.~\citep{Hilbeetal2013}.

Equalizer ZD strategies (eZDs) are selfish and mean: they force a fixed payoff (in most cases, a payoff lower than what eZD receives) on the opponent, who has no control over the matter: the fixed payoff  (in the limit of a large number of plays) is independent of the opponent's strategy vector $\vec P$.  It is not easy to see how such a strategy can be possible, but the mathematics is borne out by the simulations: in direct clashes, eZD players wins against most other strategies (but not all of them: playing All-D [always defect] is a good defense against eZD, because while All-D's payoff is fixed by eZD, it actually exceeds what eZD receives). 

However, the payoffs between opposing players do not determine the fate of populations, because in well-mixed populations of both types, agents must play their kin as often as they play a different type. Thus, a successful strategy must play nicely against their own kind, which eZD does not do because they fix the opponent's payoff {\em regardless} what their strategy is. Thus, eZD not only rips off others, it also rips off eZD, that is, itself. An agent-based simulation at zero mutation rate reveals immediately that eZD is not stable against a variety of opponents~\citep{AdamiHintze2013}. Most damning for eZD is its fate against the strategy {\sc Pavlov}, defined by the strategy vector $\vec P_{\rm Pav}=(1,0,0,1)$. Even though eZD wins each and every single encounter with {\sc Pavlov}, it will quickly be driven to extinction because  {\sc Pavlov} cooperates with its kin, while eZD treats its own kind like it treats everybody: shabbily. 

The compliant or generous ZD strategies fare much better in populations than the selfish ones, and mathematics suggests that they are in fact invincible, in the sense that they are evolutionarily stable~\citep{Akin2012,StewartPlotkin2013}. But are they really?

\subsection*{Mutational robustness and evolutionarily stable quasistrategies}
Stewart and Plotkin~\citep{StewartPlotkin2013} define a strategy to be {\em mutationally robust} if it cannot be invaded by any other strategy, except when that strategy is neutral (that is, has the same fitness), in which case it can be replaced with probability $1/N$, where $N$ is the population size. However, it was shown recently~\citep{Leeetal2015} that Stewart and Plotkin's robust and generous strategy \zdr\ (defined by the probability vector  $(1.0, 0.35, 0.75, 0.1)$ in the order defined by (\ref{stoch})) is easily outcompeted by strategies that can distinguish self from non-self. The optimal strategy of this kind is the conditional defector CD that we defined in~\citep{AdamiHintze2013}: it can invade \zdr, but it can resist invasion by \zdr~\citep{Leeetal2015}. However, CD is not a memory-one strategy, so perhaps \zdr\ is still the king of the memory-one strategies? It turns out that this is only the case in the SSWM regime discussed earlier, where the theoretical estimates of fixation can be trusted. At finite mutation rates, \zdr\ will lose against other more robust memory-one strategies, which requires us to rethink the concept of mutational robustness as defined in~\citep{StewartPlotkin2013}. 
\begin{figure}[htbp] 
   \centering
   \includegraphics[width=6.5in]{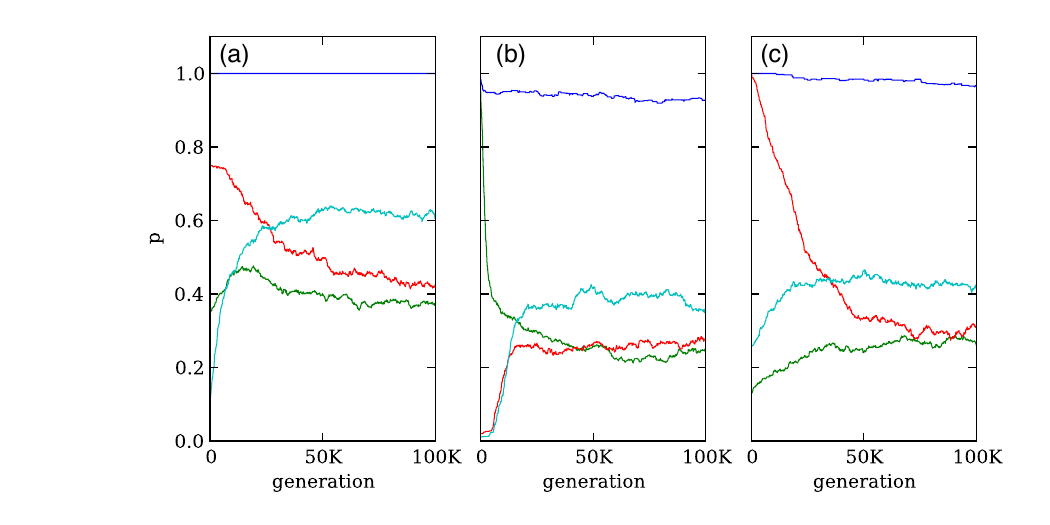} 
   \caption{Average strategy evolution along the LOD. All runs use $N=1,024$, $\mu=1\%$ per locus. We show only the first 100K of $2\times 10^6$ generations. (a) Average LOD that begins with a population of \zdr\ strategies with $\vec P_{{\rm ZD}_{\rm R}}=(1.0, 0.35, 0.75, 0.1)$.
 The strategy vector moves away from \zdr\ and ends up at the GC strategy for the donation game, with payoff matrix $(R,S,T,P)=(2,-1,3,0)$~\protect\citep{StewartPlotkin2013}. (b) Average LOD originating at the equalizer ZD strategy  $\vec P_{\rm ZD}=(0.99,0.97,0.02,0.01)$. eZD is unstable and moves towards the GC strategy $\vec P_{\rm GC}=(0.91, 0.25, 0.27, 0.38)$ for payoffs $(R,S,T,P)=(3,0,5,1)$. (c) Average LOD with ZDGTFT-2~\citep{StewartPlotkin2012} as ancestor (same payoff matrix as in (b)). ZDGTFT-2 is unstable and moves towards $\vec P_{\rm GC}$. All strategies use $p_0=0.5$, that is, they have an even chance of cooperating on the first move.
 \label{fig:mut-robust}}
\end{figure}

{\bf Stochastic conditional strategies that are stable in the SSWM regime may be unstable in the WSSM regime.} 
In Fig.~\ref{fig:mut-robust} we show the result of agent-based simulations where the population is initialized with three different resident ZD strategies, and allowed to evolve at a mutation rate of 1\% per locus, and a population size of 1,024 agents in a well-mixed setting, as in~\citep{Iliopoulosetal2010}. We repeat the experiment 200 times, and plot the average of the four probabilities defining the strategy along the evolutionary line of descent (LOD, see Box 3). The LOD recapitulates the evolutionary history of the particular experiment: it is obtained by identifying the most common strategy at the end of the run, and then identifying the chain of strategies that led to it backwards, mutation by mutation all the way to the ancestor. In this way, an unbroken line of mutated genotypes leads from the starting strategy to the most recent common ancestor (MRCA) in the population. (Because every population has multiple lines, each beginning with a different existing genotype in the population, we show as the last genotype in the line the unique MRCA, or a genotype close to it.) The LOD of each particular run is fairly jagged, but the mean LOD averaged over many runs paints a fairly detailed picture of adaptation. We find that each of the three ZD ancestors is ultimately replaced by a robust memory-one strategy that is not a ZD strategy, something that should not be possible according the mutational robustness definition described above.  What is the nature of these winning strategies?

It turns out that the winning strategies are not so much identifiable strategies given by a single strategy vector $\vec P$, but rather they are stable {\em groups of strategies} that mutate into each other and support each other that way. In that sense, they form a quasispecies~\citep{Eigen1971,EigenSchuster1979} and we refer to them as {\em quasistrategies}~\citep{AdamiHintze2016}. A quasistrategy 
is described by a probability {\em distribution} of strategies, which is a solution to an equation that is analogous to the replicator equation of EGT, but with an added mutation term. In unconditional games, the distribution is a fixed point of the stochastic replicator equation that we describe below. For conditional games, the replicator equation cannot be written down in that manner, but when mutations are present we can show that the distribution is completely determined by its mean.

Let us define an unconditional stochastic strategy ${\mathbf S}(p)=(p_1,...,p_n)$ where the $p_i$ are probabilities to engage in any of $n$ different strategies. Let $E$ be the $n\times n$ payoff matrix, while $f(p)=f(p_1,....,p_n)$ describes the population distribution of strategies. Then the population mean strategy is
\be
\bar {\mathbf S}=\sum_p f(p) {\mathbf S}(p)\;,
\ee
where we have assumed a discretization of strategy space (these averages can be generalized to continuous strategy spaces, see, e.g.,~\citep{Hines1980,Zeeman1981}). 
The distribution obeys the replicator equation
\be  \label{diff}
\frac{\dot f(p)(t)}{f(p)}=\left({\mathbf S}(p)-\bar {\mathbf S}\right)\cdot E \bar {\mathbf S} \;.
\ee
In the presence of mutations, the distribution becomes stationary $\frac d{dt}f(p)=0$ and the population mean strategy is a solution to $\left({\mathbf S}(p)-\bar {\mathbf S}\right)\cdot E \bar {\mathbf S}=0$, given by a fixed point on the support of $f(p)$ (an $n-1$-dimensional simplex). Depending on the payoff matrix, the fixed point lies either in the interior of the simplex (in which case all $S_i$ are between zero and 1), or they can lie on one of the faces. In any case, the population fixed point is given by the corresponding fixed point of the deterministic game~\citep{Zeeman1981}. For conditional games, however, we cannot write down the replicator equation because there is no payoff matrix that can be written down, and the support of the probability distribution is an $n$-simplex (as the sump of the $p_1$ need not equal 1)~\citep{AdamiHintze2016}. However, if we assume that the distributions are stationary, it is possible to relate the population fixed point $\bar {\mathbf S}$ to the distribution $f(p)$ as follows. 

\begin{figure}[htbp] 
   \centering
   \includegraphics[width=3in]{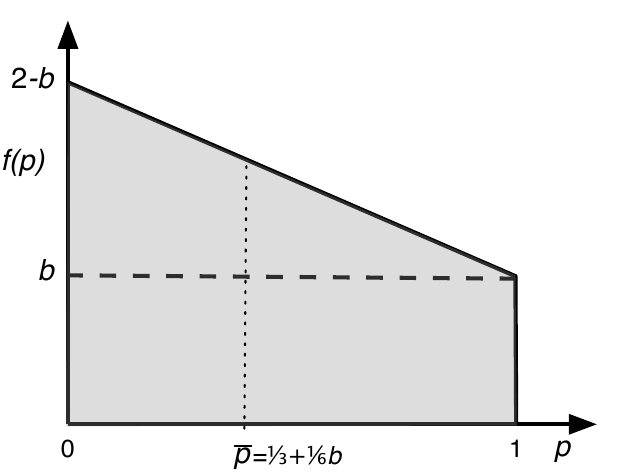} 
   \caption{Probability distribution for a stochastic strategy determined by a single probability $p$. The mean $\bar p$ of the distribution $f(p)$  coincides with the corresponding population mixture ESS of the deterministic game, and is determined entirely by one of the boundary values of the distribution (here $b$).}
   \label{fig:diff1}
\end{figure}

\begin{figure}[htbp] 
   \centering
   \includegraphics[width=4in]{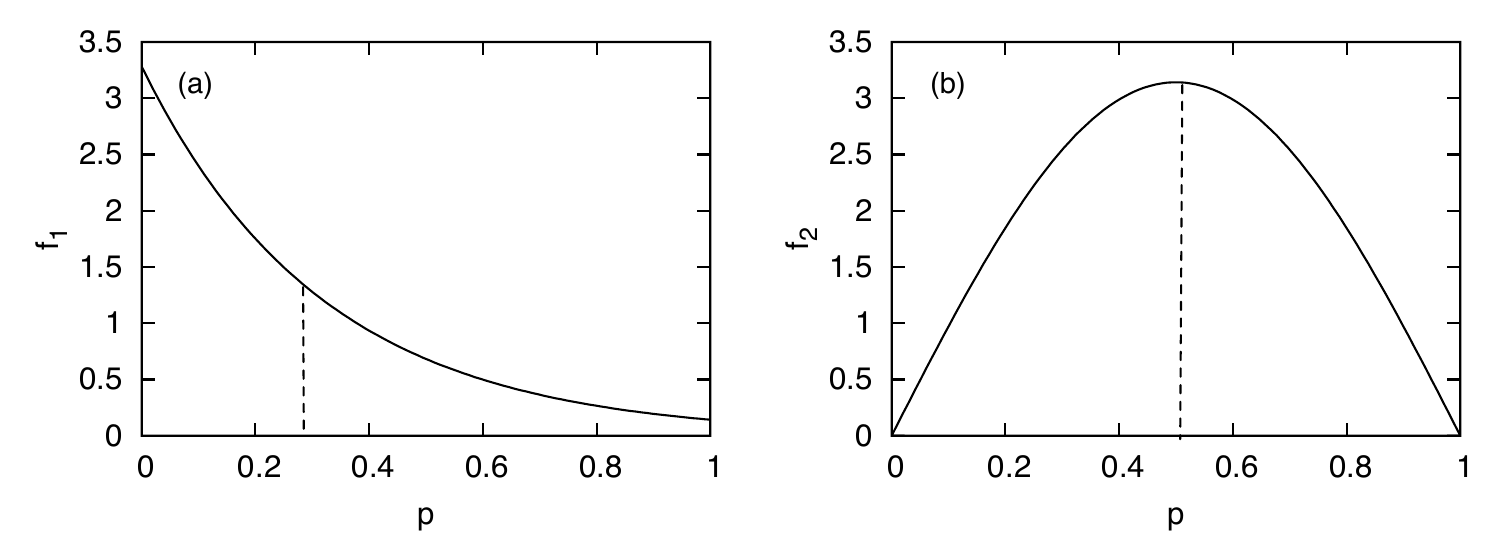} 
   \caption{Marginal probability distributions $f_1(p_1)$ and $f_2(p_2)$ of a stochastic game with $n=3$. Here the means $\bar p_1\approx0.27$ and $\bar p_2=0.5$ are determined by the boundary conditions $f_1(1)\approx0$ and $f_2(1)=0$.}
   \label{fig:diff2}
\end{figure}

\begin{figure}[htbp] 
   \centering
   \includegraphics[width=4.5in]{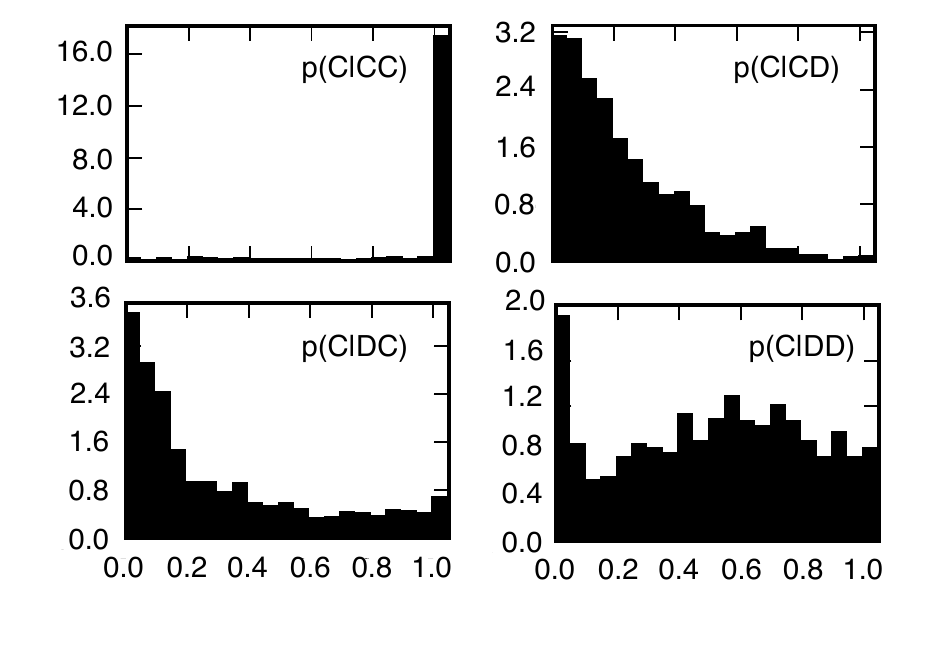} 
   \caption{Marginal distribution functions $f(p_i)$ of probabilistic conditional (memory-one) strategies in an adapted population ($N=1,024, \mu=0.5\%$) after 1.75 million generations, using the payoffs of the donation game, 200 replicate populations. The mean strategy is $\bar {\mathbf S}=(1.0,0.27,0.3,0.6)$. The distribution of strategies along the LOD shown here recapitulates the population distribution (not shown), suggesting that the dynamics are ergodic.}
   \label{fig:diff3}
\end{figure}

If mutations are local and homogenous (they occur in the same manner on all probabilities), then the distribution $f(p)$ must obey the diffusion equation~\citep{Zeeman1981}. If the distribution is stationary, this is Laplace's equation
\be
\sum_{i=1}^n\frac {\partial^2 f(p)} {\partial p_i^2}=0\;.
\ee   
That the boundary conditions (the values of $f$ at $p_i=0,1$) can determine the population mean is not immediately obvious, but is best illustrated with the case $n=2$, for which there is only one probability $p$ (the other being $1-p$). In that case, the Laplace equation implies that the distribution must be linear, and if the distribution is normalized the mean of that distribution is completely determined by one boundary value. For example, if the distribution is $f(p)=a+(b-a)p$ with $a=2-b$ for normalization, the population mean is $\bar p=1/3+b/6$ (see Fig.~\ref{fig:diff1}), that is, it is entirely determined by one of the two boundary values. Put in another way, the mean of the distribution unambiguously determines the boundary value, and thus there is a one-to-one correspondence between means and probability distributions, but note that the boundaries themselves could deviate from the distribution, as noted by Zeeman~\citep{Zeeman1981}. Furthermore, the mean is not anymore determined by the fixed point of the deterministic game, but will change with the mutation rate. 
In more dimensions, the solution to the Laplace equation is less straightforward. For two independent probabilities $p_1$ and $p_2$, the diffusion equation can be solved if we can assume that the variables can be separated as $f(p_1,p_2)=f_1(p_1)f_2(p_2)$ (something that is not always warranted as we will discuss below). Then
\be
\frac1{f_1}\frac {\partial^2 f_1(p_1)} {\partial p_1^2}+\frac1{f_2}\frac {\partial^2 f_2(p_2)} {\partial p_2^2}=0\;,
\ee 
which can only be satisfied if 
\be
\frac {\partial^2 f_1(p_1)} {\partial p_1^2}&=&\lambda  f_1(p_1) \;, \\
\frac {\partial^2 f_2(p_2)} {\partial p_2^2}&=&-\lambda  f_2(p_2)
\ee
separately.
A typical solution where the $f_i$ are positive and where $f_1(1)\approx0$ and $f_2(1)=0$ is 
\be
f_1(p_1)&=&A_1(\cosh(\pi p_1)-\sinh(\pi p_1))\;,\\
f_2(p_2)&=&A_2\sin(\pi p_2)
\ee
with suitable normalization coefficients $A_i$ that do not affect the mean. With these boundary conditions, the means are uniquely specified (see Fig.~\ref{fig:diff2}).
We show in Fig.~\ref{fig:diff3} the distributions $f_i(p_i)$ for the four conditional probabilities $p_1,p_2,p_3,p_4$ defined in (\ref{stoch}), which show shapes remarkably similar to the theoretical shapes shown in Fig.~\ref{fig:diff2}). Indeed, the distribution $f_1(p_1)$ ``decouples", meaning that as it is fixed to the boundary, the population fixed point occurs on the remaining three-dimensional subspace. The solution to the Laplace equations are two hyperbolic and one trigonometric function as is easily checked.
Earlier we made the assumption that the population probability distribution $f(p)$ factorizes, so as to be able to solve the Laplace equation. Writing $f(p)$ in this way is always possible mathematically, but whether or not the marginal distributions correctly describe the joint distribution is not immediately given. 
We tested the pairwise correlation coefficients for the distributions $f_2$, $f_3$ and $f_4$ (as $f_1$ shows virtually no variance, it cannot be correlated to any of the other distributions) and found correlation coefficients smaller than 0.05 throughout. Thus, we can be confident that the marginal distributions (and their means) describe the population strategy accurately. 

That the distribution of strategies $f(p)$ is stationary is a hallmark of the quasispecies concept, which is the stationary solution to the replicator-mutator equations advanced by Eigen and Schuster~\citep{EigenSchuster1979}. The quasispecies naturally depends on the mutation rate, and in particular on the shape of the fitness peak. It is possible, for example, that a species occupying a lower fitness peak that is broad (and thus exhibits more mutational robustness) can outcompete a species occupying a higher peak that is narrower~\citep{SchusterSwetina1988,Wilkeetal2001,WilkeAdami2003}, a concept commonly knows as ``survival of the flattest". The theory also implies that the population becomes random (and the distribution function uniform) once a threshold mutation rate is reached (the ``error threshold"), and such a transition is readily observed~\citep{Iliopoulosetal2010}.

In Fig~\ref{fig:mut-robust}a, the ancestor is the generous ZD strategy \zdr\ that according to mathematics~\citep{StewartPlotkin2013} cannot be invaded by any other memory-one strategy (except neutrally). However, it is clearly replaced by a quasistrategy almost immediately, and in particular the not-very-generous probability to cooperate after DD plays ($p_4=0.1$) moves away from this value towards a much more forgiving value. The quasistrategy that is ultimately selected is a robust cooperating strategy that we term ``General Cooperator" (GC). It is (at $\mu=1\%$) determined by the mean $\vec P_{\rm GC}=(1.0, 0.36\pm0.02, 0.42\pm0.01, 0.64\pm0.01)$~\footnote{Note that the mutual cooperation probability remains at $p_{CC}=1$ and does not drift. In the following, we refer to the GC strategy with the donation game payoff matrix as ${\rm GC}_d$.}  (the probabilities quoted were obtained using generations $1\times 10^6$ to $1.95\times 10^6$ of the LOD, averaged over 200 replicate runs, showing 95\% confidence intervals). 


Note also that the mean probabilities defining this particular GC strategy are not the same as the one obtained at the same mutation rate earlier~\citep{Iliopoulosetal2010,AdamiHintze2013}, because the payoff matrix for \zdr\ is from the donation game rather than the standard (R,S,T,P)=(3,0,5,1). Note further that a pure GC cannot invade \zdr (starting a population with a single strategy encoding the means of the quasi-strategy), and loses a direct matchup. However, as the quasistrategy forms, the resident strategy \zdr\ is outcompeted because it loses (on average) in a direct competition against its own mutants, which in turn are themselves unstable as we will show below. We tested explicitly that \zdr\ only becomes stable for mutation rates smaller than $1/N^2$. 

In Fig~\ref{fig:mut-robust}b, the ancestor is a selfish equalizer ZD strategy of Press and Dyson~\citep{PressDyson2012}: its evolutionary demise was already observed in~\citep{AdamiHintze2013} and is therefore not a surprise.  The evolutionary fixed point is the GC quasistrategy of~\citep{Iliopoulosetal2010}, a generous quasi-strategy with $\bar p_4\approx 0.4$, which does not cooperate fully after CC so as to be able to take advantage of $p_1=1$ strategies. We call it ${\rm GC}_A$ as it is obtained using the Axelrod payoffs. Running the simulation to $2\times 10^6$ generations as before, we obtain the more accurate probabilities
\be
\vec P_{{\rm GC}_A}=(0.91\pm0.02, 0.25\pm0.02, 0.27 \pm 0.03, 0.38\pm0.03)\;.
\ee
Because the temptation payoff $T$ is higher in this game than in the donation game payoff, occasionally ripping off cooperators pays off and $p_{\rm CC}<1$.

Another generous ZD strategy is ZDGTFT-2, a generous version of TfT with $\vec P_{\rm ZDGTFT-2}=(1,1/8,1,1/4)$ that is in the ZD class, introduced in~\citep{StewartPlotkin2012}. It is similarly unstable in the WSSM regime, as we can see in Fig.~\ref{fig:mut-robust}c. The evolutionary instability of \zdr\ and ZDGTFT-2, which both do exceedingly well in the SSWM regime, can be traced back to how they play against their own mutants. To see this explicitly, we discretized the space of all stochastic memory-one strategies using probabilities 
\be
p_i^{(n)}=\frac{n}{20}\;, \ \ \ \ \ i=1,...,4;\ \ \  n=0,...,20\;,
\ee
and added to this space the probability of the originating type (because that is unlikely to be contained in the discretization) giving rise to a strategy space of between $21^4=160,000$ up to $25^4=361,776$ strategies. All strategies play C or D with probability $p_0=0.5$ on the first move. We then played each tested strategy $S$ against {\em all} of its one-mutants (the set $S(1)$) for 500 iterations each\footnote{The average payoff after 500 iterations is indistinguishable from the exact result (infinite number of iterations) obtained by diagonalizing the corresponding Markov matrix.}, and recorded the score $E[S,S(1)]$ (the average payoff of the strategy against its one-mutant), what the one-mutants receive against the strategy, $E[S(1),S]$, and what the average mutant receives against itself, i.e., $E[S(1),S(1)]$.
For the \zdr\ strategy on the one hand we calculate an effective payoff matrix 
\be
\bordermatrix{\mbox{} & {\rm ZD}_{\rm R} & {\rm ZD}_{\rm R}(1)  \cr
                           {\rm ZD}_{\rm R}     & 2.0  & 1.7 \cr
                            {\rm ZD}_{\rm R}(1)     & 1.77 &  1.66   }\;.  \label{zdr-pay}
\ee  
We note that while \zdr\ is an ESS with respect to its one-mutants (as $2.0>1.77$), the direct competition goes in favor of the mutants ($1.77>1.7$). At the same time, the mutants themselves are not stable ($1.66<1.7$). 
 
For the mean ${\rm GC}_d$ strategy that \zdr\  evolves into, on the other hand, we obtain 
\be
\bordermatrix{\mbox{} & {\rm GC}_{d} & {\rm GC}_d(1)  \cr
                           {\rm  GC}_d      & 2.0  & 1.847 \cr
                            {\rm GC}_d(1)      &1.92 &  1.851   }\;. \label{GCd-pay}
\ee 
The mean strategy ${\rm  GC}_d$ is also ESS against its own one-mutants, and it also loses against its mutants in a direct matchup. However, the mutants are stable ($1.851>1.847$), and can therefore coexist with the (mean) wild-type. At the same time, the fitness of the mutants (measured in terms of the payoff against its own kind) is significantly higher than the \zdr\ mutant fitness (1.851 compared to 1.66). It is in this manner that the quasistrategy achieves its mutational robustness, and a stable distribution in the population.

\begin{figure}[htbp] 
   \centering
   \includegraphics[width=4in]{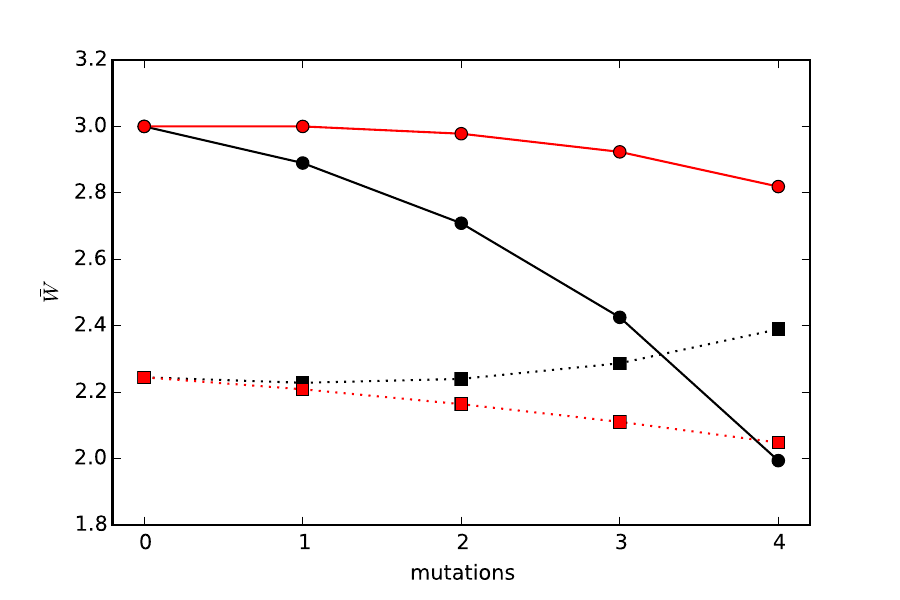} 
   \caption{Mean payoff to a strategy from its mutants, compared to the payoff received by the mutants. Average payoff $E[S,S(i)]$ for $S$=gTFT (black solid circles) against all of its $i$-mutants $S(i)$, and payoff of the $i$-mutants $E[S(i),S]$ (red solid circles). The payoffs for the strategy that gTFT evolves into are given by the black and red solid squares, with the black squares (payoffs to resident GC wild-type) always larger than the red squares (payoff to mutants from GC).}
   \label{mut-plot}
\end{figure}

To test if this trend holds for other generous ZD strategies, we repeated the analysis for the strategy ZDGTFT-2~\citep{StewartPlotkin2012} (abbreviated ${\rm ZD_2}$), to find
\be
\bordermatrix{\mbox{} & {\rm ZD}_2 & {\rm ZD}_2(1)  \cr
                           {\rm ZD}_2     & 3.0  & 2.93 \cr
                   {\rm ZD}_2(1)     & 2.96 &  2.91   }\;,  \label{zdgtft-pay}
\ee  
compared to the GC strategy evolved with Axelrod's payoffs ${\rm GC}_A$ that ${\rm ZD}_2$ evolves into:
\be
\bordermatrix{\mbox{} & {\rm GC}_A & {\rm GC}_A(1)  \cr
                           {\rm  GC}_A     & 2.24  & 2.23 \cr
                            {\rm GC}_A(1)      &2.21 &  2.24   }\;.  \label{GCA-pay}
\ee 
We find a similar picture: ${\rm GC}$ one-mutants are ESS (as is the wild-type against the mutants), but in this case the mutants actually lose against the wild-type in a head-to-head competition. At the same time, the fitness differential between the wild-type and its mutants is extremely small. From a fitness landscape point-of-view, we deduce that GC occupies a flat fitness peak where the peak and its mutants are effectively neutral, forming a connected set of Nash equilibria akin to the ES-sets discussed above. It has been shown previously that more connected neutral networks can outcompete less connected ones~\citep{Nimwegenetal1999,Wilke2001}, and even that types that live on flat peaks can outcompete types with higher fitness (but occupying a narrower peak) as long as the mutation rate is high enough: the ``survival of the flattest" effect~\citep{Wilkeetal2001}. The flat structure of the peak makes it possible to establish a stationary mutant distribution that confers stability to the wild-type (this is akin to the formation of a quasispecies~\citep{Eigen1971,EigenSchuster1979,SchusterSwetina1988}). 
This is seen most easily when we plot the average payoff to the strategy against its $i$ mutants
 (with $i=1\cdots4$) $E[S,S(i)]$, as well as the payoff that the strategy mutants $S(i)$ receive against the wild-type strategy, $E[S(i),S]$. We show this in Fig.~\ref{mut-plot} for the resident ZD strategy gTFT (generous TfT), given by the probabilities (for the Axelrod payoffs) 
 $\vec P_{\rm gTFT}=(1,1/3,1,1/3)$.
The ``nice" ZD strategy gTFT receives a poor payoff against its one-mutants (``does not play well with its mutants"), while essentially all mutants of gTFT score close to maximum (red circles in Fig.~\ref{mut-plot}). While the gTFT payoffs are significantly larger than what GC receives against itself (and against its own one-mutants), the difference in mutational stability comes from the fact that GC obtains more than its mutants in a head-to-head competition (black curve above red curve in Fig.~\ref{mut-plot}), as well as from the stability of the mutant distribution. This inversion together with the closeness of the red and black curves creates the flat fitness landscape that ultimately leads to the evolutionary demise of the resident ZD strategy. If the mutation rate is sufficiently low ($\mu<1/N^2$) we expect gTFT to be stable as it is a Nash equilibrium against its own one-mutants, and at that mutation rate the chance that two mutants exist in the population(alongside with the resident strategy) is very low. 

We thus conclude that evolutionary and mutational stability in the WSSM regime (which is the biologically relevant regime, and also the regime used in most standard Genetic Algorithms) is different from evolutionary stability in the SSWM regime that is mathematically tractable. Indeed, the dynamics of strategies in the WSSM regime is reminiscent of the dynamics of quasispecies, where groups of genotypes that live on low fitness peaks that are flat (have a large fraction of neutral neighbors) can outcompete types living on higher fitness peaks but with a steep drop, as long as the mutation rate is sufficiently high. We thus find that the conventional concept of the ESS is, in this regime, replaced by a mutation rate-dependent evolutionarily stable quasistrategy~\citep{AdamiHintze2016}. It is well-known that the full quasispecies equation can only be solved analytically for very special fitness landscapes~\citep{Wilke2005}, so agent-based simulations are key to exploring this new domain for evolutionary games.

\subsection*{Multi-player games}
Multi-player games can be seen as an extension of the games we just discussed, where payoffs are given to an agent and its opponent. In multi-player games, the payoff is a function of the state of more than two players in a group. The standard game in this category is the Public Goods (PG) game, which reduces to the PD for two players. The payoffs for this game are not usually written in terms of a payoff matrix (as such a matrix assumes only pair-wise interactions), but we will see that in the limit of the well-mixed game, the fixed points of the replicator equation can be obtained using a payoff matrix after all. 

The Public Goods game is a staple of experimental economics~\citep{Olson1971,DavisHolt1993,Ledyard1995} where players can invest tokens  into a common pool (the public good). The contributed sum is then multiplied by a ``synergy factor" that amplifies the public good that is subsequently equally distributed to the players, irrespective of whether they paid in or not. Groups of players maximize their investment if 
everybody in the group contributes, so as to take maximum advantage of the amplification. This behavior is vulnerable to ``free-riders" that share in the re-distribution of the public good, but do not invest themselves. It is easy to show that the rational Nash equilibrium for this game is to defect, because this strategy clearly dominates all others regardless of their play~\citep{Hardin1968}.

We will study a game with $k+1$ players in a group (the $k$ participants plus the ``focal" player), with a linear synergy function with factor $r$. If we denote with 
$N_C$  the number of cooperators among the participants (not counting the focal player so that $N_C\leq k$) and $N_D$  the number of defectors, the net payoff a cooperator receives (after subtracting the donation) is
\be
P_C = r\frac{(N_C+1)}{k+1} -1 \label{eq1a}\;,
\ee
while the defector gets
\be
P_D=r\frac{N_C}{k+1}\;. \label{eq2a}
\ee
If it is advantageous for an individual to defect while at the same time mutual cooperation would be beneficial for all, a dilemma exists. A defector does better as long as $P_D-P_C>0$, so the dilemma exists only if $r<k+1$. Also, it is clear that the payoff for a cooperator playing within a group of cooperators should exceed the payoff to a defector surrounded only with defectors, which implies $P_C(N_C=k)-P_D(N_C=0)>0$ or $r>1$. Thus, a dilemma exists in this game for the range of parameters for $1<r<k+1$.

\begin{figure}[htbp] 
   \centering
   \includegraphics[width=3in]{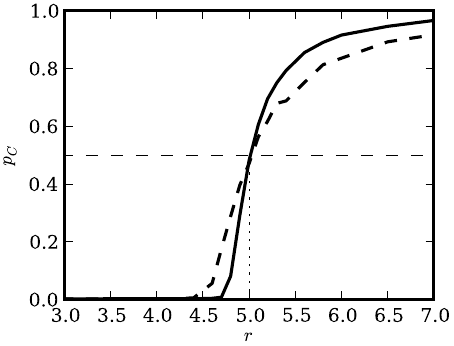} 
   \caption{{\bf Mean probability of cooperation in PG game}. Probability of cooperation $p_C$ as a function of the synergy $r$ for $k=4$, $N=1,024$, $\mu=2\%$ (dashed) and $\mu=1\%$ (solid). Average of 100 replicate runs, 250K updates from each run lasting 500K updates, discarding the first 100K updates (the transient) and the last 50K.  Modified from~\citep{HintzeAdami2015}.}
   \label{fig:pg0}
\end{figure}

Solving the replicator equation for this game is straightforward and shows that defection is the evolutionary stable strategy for $1<r<k+1$, while cooperation is stable for $r>k+1$. This can be seen most easily by writing an effective payoff matrix for this game, in which we write the payoff received by a single cooperator against a uniform background of other cooperators $E(C,C)=r-1$ (setting $N_C=k$ in (\ref{eq1a})), or in a background of only defectors $E(C,D)= \frac{r}{k+1} -1$ ($N_C=0$ in  (\ref{eq1a})), and similarly for defectors, to arrive at the ``invasion matrix"
\be
\bordermatrix{\mbox{} & {\rm C} & {\rm D}  \cr
                           {\rm  C}      & r-1  & \frac{r}{k+1} -1 \cr
                            {\rm D}      &\frac{rk}{k+1} &  0   }\;.  \label{PD1}
\ee  
This matrix can be written into normal form by subtracting $r-1$ from the cooperator column as
\be
\bordermatrix{\mbox{} & {\rm C} & {\rm D}  \cr
                           {\rm  C}      & 0  & \frac{r}{k+1} -1 \cr
                            {\rm D}      &-\frac{r}{k+1} +1 &  0   }\;.  \label{PD2}
\ee  
This form immediately confirms that the deterministic form of this game has one attracting and one repulsive fixed point (cf. Box 1), with a critical point that flips between the two at $r=k+1$. Agent-based simulations of the evolution of {\em probabilistic} strategies that encode a single locus (the probability to cooperate) confirm this prediction, as seen in Fig.~\ref{fig:pg0}. There, we show the average probability to cooperate (obtained by averaging the mean probability to cooperate on the stable part of the evolutionary LOD up to the MRCA) as a function of the synergy parameter $r$, obtained with two different mutation rates in a population of $N=1,024$ agents, with $k=4$ and 100 replicates per $r$, for two different mutation rates~\citep{HintzeAdami2015}. 

It is clear that we can view the transition from defection to cooperation (as $r$ is increased) as a phase transition, with coexistence of phases at the critical point $r=k+1$, but where the transition towards ``order" (if we would name the cooperating phase thus) is initated quite a bit before the critical point. The continuous nature of the transition is a standard feature due to the finite population size, but what is less expected (and cannot be deduced from the analytical treatment), is that the shape of the transition depends on the mutation rate. The agent-based simulations suggest that the transition is a broadened first-other transition, where the broadening via mutations is akin to the broadening observed in physical systems due to random quenched impurities~\citep{ImryWortis1979}. We hasten to add that in this probabilistic simulation, there is only one phase (the strategy playing with probability $p_C$), but as the transition recapitulates what we would see in a deterministic implementation (in which we expect a rounder first-order transition), we keep the imagery.

Given that the dilemma prevents cooperation, can we use communication (as in the PD) to ensure cooperation? It turns out that we can, by applying the formalism of ZD strategies to the PG game~\citep{Hilbeetal2014,Hilbeetal2015,Panetal2015}\footnote{In fact, Hilbe et al.\ show that there are ZD strategies in any game with a social dilemma~\citep{Hilbeetal2015}.}.
To see this, we can study conditional strategies where the probability to cooperate depends on the number of players that have cooperated in the previous round. For a game with $k$ neighbors, the probability vector of a memory-one strategy in the PG game (also studied numerically by Hauert and Schuster~\citep{HauertSchuster1997}) can be written as~\citep{Hilbeetal2014}
\be
\vec P=(p_{C,k},p_{C,k-1},...,p_{C,1},p_{C,0};p_{D,k},p_{D,k-1},...,p_{D,1},p_{D,0})\;,
\ee
where $p_{S,j}$ is the probability of the focal player to cooperate if he played $S\in\{C,D\}$ in the previous round, and $j$ of his opponents cooperated at the same time. Among the ZD strategies that can be defined within the class of memory-one conditional PG strategies, it is possible to find exploitative and generous ones, just as in the PD game (which after all is the $k=1$ limit of the PG game). One of the fair ZD strategies is a version of the TfT strategy called ``proportional TfT" (pTfT), which cooperates with a probability given by the density of cooperators in the previous round (irrespective of the own previous move)
\be
\vec P=(1,\frac{k}{k+1},...,\frac1{k+1};\frac{k}{k+1},...,\frac1{k+1},0)\;.
\ee
The strategy pTFT can achieve a mean sustained payoff of $(r-1)/2$, but it is not clear whether evolution would favor that strategy. Hilbe et al.\ performed evolutionary simulations in the SSWM regime~\citep{Hilbeetal2015}, suggesting that while fair and generous communicating ZD strategies can elevate the mean payoff of a population in the dilemma region, cooperation is not stable in larger groups (larger $k$), as the increased group size exacerbates the temptation to defect if $r$ is constant. 

Whether this conclusion also holds in the WSSM regime is an open question, but we can ask whether there are other forms of communication that can establish cooperation in the dilemma region. For the PD, for example, the most successful strategy is the conditional defector CD~\citep{AdamiHintze2013,Leeetal2015}. That strategy is not a memory-one strategy, but rather one that can distinguish self from others with perfect accuracy, that is, it can use one bit of information (if we assume an uninformative prior) to its advantage. If it is possible for a strategy to distinguish between cooperators and defectors, then such a strategy can also {\em punish} defectors in an attempt to suppress them.

Punishment was introduced into the PG game to study whether punishment alone can establish cooperation in the dilemma region $1< r< k+1$. In that region,  defectors prevail even though cooperation would be beneficial for all. The game with punishment has been studied extensively (usually without realizing that punishment actually entails communication), both theoretically and in agent-based simulations~\citep{Yamagishi1986,Sigmundetal2001,SzaboHauert2002,FehrGachter2002,FehrFischbacher2003,Hammerstein2003,Fowler2005,NakamaruIwasa2006,CamererFehr2006,Gurerketal2006,Hauertetal2007,DeSilvaetal2009,HenrichBoyd2001,Boydetal2003,Brandtetal2003,Helbingetal2010,Helbingetal2010c,Boydetal2010,Sigmundetal2010,Sasakietal2011,Szolnokietal2011,PercSzolnoki2012,SzolnokiPerc2013a,SzolnokiPerc2013b,Chenetal2014,HintzeAdami2015}. This work shows that punishment is a two-edged sword: While punishment can be effective, it is also costly, and gives rise to another dilemma: the curse of the ``second-order free rider": cooperators that choose not to punish. 

In the game with punishment, four possible strategies emerge: cooperators that punish defectors (we call them ``moralists" after~\citep{Helbingetal2010}), ``naked" cooperators that do not punish, the usual defectors, and then the awkward defectors that punish other defectors, aptly named ``immoralists" by Helbing et al.~\citep{Helbingetal2010}. We can calculate the fixed point of the dynamics with punishment in the well-mixed regime by defining the mean density of cooperators $\rho_C$, as well as the mean density of punishers $\rho_P$ (here, $N_M$ is the number of moralist players, and $N_I$ the number of immoralists)
\be
\rho_C&=&\frac{N_C+N_M}k \label{rhoc}\\
\rho_P&=&\frac{N_M+N_I}k \label{rhop}\;.
\ee
If we also define the mean density of defectors as $\rho_D=\frac{N_D+N_I}k $, we can write the mean payoffs for 
the four possible strategies as
\be
P_C &=& r\frac{(k\rho_C+1)}{k+1} -1\;, \label{eq1c} \\
P_D&=&r\frac{k\rho_C}{k+1}-\beta \rho_P\;,\label{eq2c} \\
P_M&=&P_C-\gamma\rho_D\;, \label{eq3c}\\
P_I&=&P_D-\gamma\rho_D\;.\label{eq4c}
\ee
In these equations, $\beta$ is the effect, and $\gamma$ the cost, of punishment. 

\begin{figure}[htbp] 
   \centering
   \includegraphics[width=\textwidth]{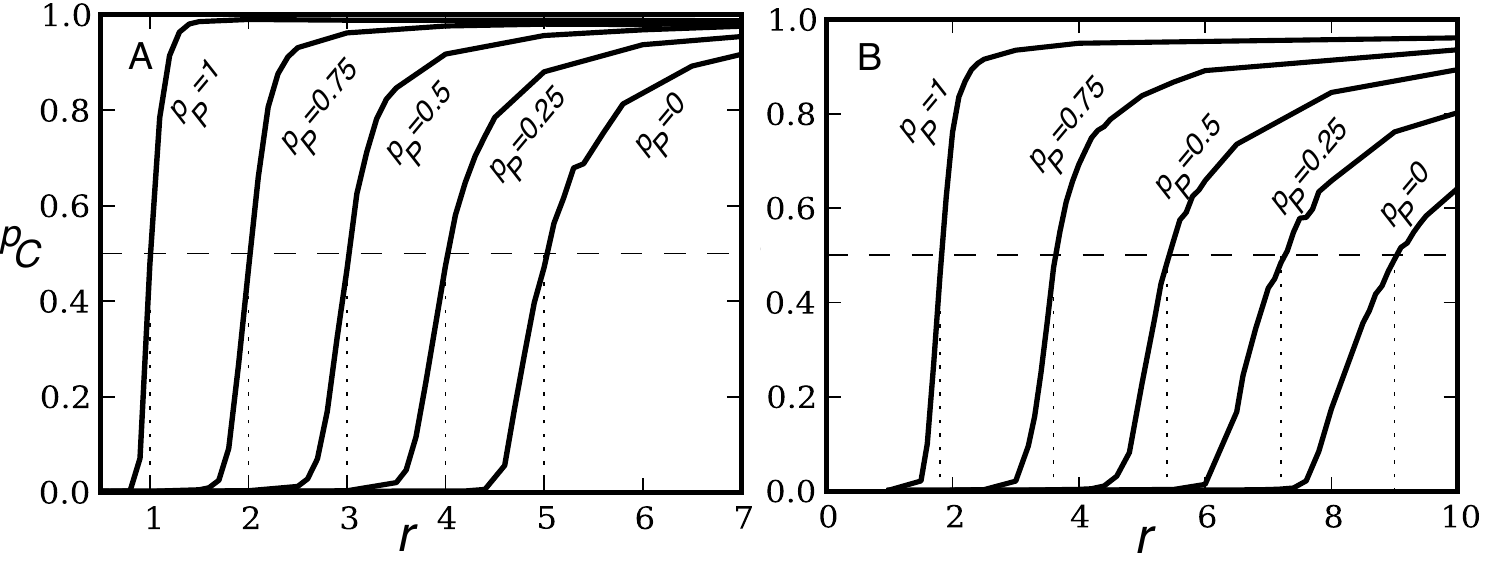} 
   \caption{Probability to cooperate, averaged over 100 independent lines of descent (average over 200K updates, discarding the first 250K and the last 50K), as a function of synergy $r$ for fixed ( that is, unevolvable) probability of punishment $p_P$=0.0, 0.25, 0.5, 0.75, and 1.0. The horizontal dashed line is at the mean probability to cooperate $p_C=0.5$, defines the critical point $r_c$. $N=1,024, \beta=0.8, \gamma=0.2, \mu=1\%$ A: $k=4$. B: $k=8$. }
   \label{fig:punish}
\end{figure}

Let us investigate the range of parameters that enable cooperation by calculating $P_C-P_D$~\citep{HintzeAdami2015}. We notice that the range where the dilemma exists is now shifted by $\beta\rho_P$ 
\be
1-\beta\rho_P<r<(k+1)(1-\beta\rho_P)\;. \label{pun}
\ee
The right boundary of the dilemma corresponds to the critical point $r_c=(k+1)(1-\beta\rho_P)$, indicating that punishers can potentially push the critical point significantly below $k+1$, in theory all the way to $r=(k+1)(1-\beta)$. This is borne out by simulations of the evolution of probabilistic strategies where the punishment probability is held fixed so that the density of punishers $\rho_P\approx p_P$, as we can see for $k=4$ in Fig.~\ref{fig:punish}A, as well as for $k=8$ (Fig.~\ref{fig:punish}B). Note that the critical point is precisely as predicted from Eq.~(\ref{pun}). 
However, the dynamics of punishment is more complex when punishment itself can evolve, that is if we encode the strategies with two independent loci (one for each probability). When defection is in full swing, punishment cannot occur as only cooperators punish (this assumes that immoralists do not play in a role in the dynamics, which is borne out in simulations). In the cooperating phase, on the other hand, punishment is unnecessary, and the gene for punishment will drift because it is not under selection (the mean probability of punishment will approach 0.5, the mean of a uniformly distributed random number bounded by zero and one). Thus, even though punishment changes the location of the critical point, it can only play a significant role when the synergy parameter happens to be close to that point.  

Punishment leads to another effect that is not easily recognized from the mean-field theory that predicts the location of the critical point: both the cooperating and the defecting phase are meta-stable near the critical point, and can collapse into the other phase if a sufficient number of the invading strategy are present. Statistical physics teaches us that this meta-stability should give rise to the phenomenon of {\em hysteresis}, where a slow change of the critical parameter past the critical point leaves the phase in a ``false state". In such cases, the phase transition is driven by nucleation events (see, e.g.,~\citep{LandauLifshitz1969}). 

\begin{figure}[htbp] 
   \centering
   \includegraphics[width=4in]{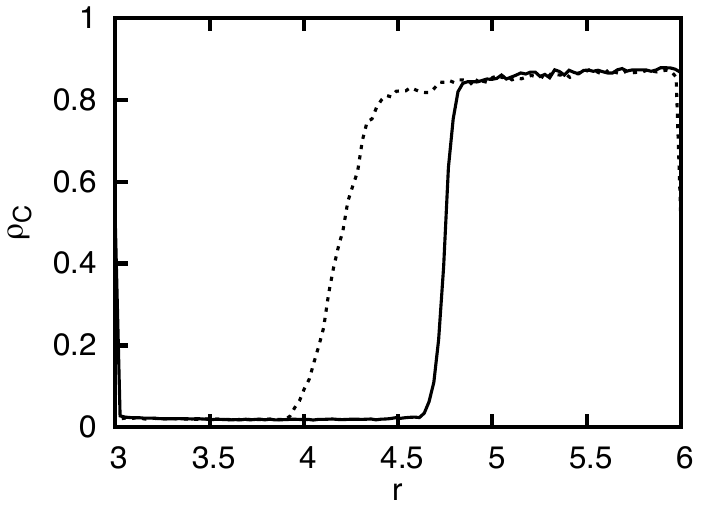} 
   \caption{{\bf Hysteresis effect from punishment.} Density of cooperators (non-punishing cooperators plus moralists) as a function of synergy $r$ when $r$ is adiabatically changed from low to high values (solid), and back from high values to low values (dashed). Here punishment can evolve, with $\beta=0.8$ and $\gamma=0.2$, $\mu=1\%$. All population fractions are started at $0.5$ (either at the high or low end of $r$). The lines show the average over 100 runs. Standard error is of the size of the fluctuations. Modified from~\citep{HintzeAdami2015}.}
   \label{fig:hyst}
\end{figure}

We conducted such a slow adiabatic changer in $r$ and found this hysteresis phenomenon in the PG game with punishment, as seen in Fig.~\ref{fig:hyst}. As the synergy $r$ is increased adiabatically (solid line), the transition occurs around $r_C\approx 4.7$ (note that punishment is free to evolve in these simulations). Decreasing $r$ from above (dashed line), the population continues to cooperate significantly past the critical point, stays in the ``false" state and crosses 0.5 (even probability to cooperate) around $r_c=4.2$ instead. This effect is essentially absent when there is no punishment. Phenomena such as hysteresis are amenable to a mathematical analysis in condensed matter physics, but it is doubtful that the observed effect can be studied analytically in game theory. Even though simple games (such as the PD and the PG in one dimension, where each player has only two neighbors) can be described in terms of an Ising-type model where punishment plays the role of a magnetic field~\citep{Adamietal2015}, it is doubtful that such a description can be extended to higher dimensions (more players in a  group), as even the 2D Ising model with a magnetic field has so far defied a closed form solution. 

\subsubsection*{Spatial interactions}
If the interaction between players are local rather than global--meaning that players interact with their immediate neighbors in space (or more generally, on a lattice) rather than with randomly chosen opponents, agent-based methods are almost unavoidable. The importance of spatial assortment is immediately obvious. After all, the reason why cooperation is not a stable strategy in the Prisoner's Dilemma  is because cooperators are vulnerable to defectors: when unconditional cooperators interact with defectors, they are ``ripped off". But if cooperators (in the absence of mutations) give rise to other cooperators that are placed next to them, then the cooperators interact with defectors only rarely, namely they encounter them only on the boundary that separates groups of cooperators from groups of defectors. Because cooperators that play with cooperators have a higher fitness than defectors that interact with other defectors, spatial (as opposed to well-mixed) dynamics is much more conducive to the establishment of cooperation~\citep{NowakMay1992,NowakMay1993,Nowaketal1994}.  Note that this is not a universal feature of all games: in the Snowdrift game for example ($a,b>0$ in Eq.~(\ref{game})), spatial interactions can suppress cooperation~\citep{HauertDoebeli2004}. Methods exist to predict levels of cooperation mathematically in spatial games via generalized mean-field calculations~\citep{SzaboToke1998,SzaboFath2007,Szolnokietal2009} or ``pair-wise methods"~\citep{HauertDoebeli2004}, but they are limited to games with few (usually only two) strategies, and are only approximations.

When strategies are unconditional, the dynamics unfolding on a grid can be ``solved" using finite-state-automaton updating of rules, akin to cellular automata~\citep{Wolfram1984}. While not a closed-form solution, these dynamics are nevertheless much simpler than the agent-based simulations that we advocate here. However, such rule-based updating schemes are impossible if the number of strategies is not limited. 
Spatial dynamics are important to understand the evolution of cooperation, as the assortment between cooperative strategies due to the placement of offspring near to the progenitor can take the place of assortment via communication (where strategies direct cooperation only towards strategies they have determined to be other cooperating strategies~\citep{Leeetal2015,Mirmomeni2015}.)  Analytical results for the spatial distribution can only be obtained in the simplest of models, for example the May-Leonard model that describes the competition between three species, akin to the Rock-Paper-Scissors game studied above. Writing the rate equations between the three species as stochastic partial differential equations can recover the spiral patterns that are observed in simulations~\citep{Reichenbachetal2007}, but it appears hopeless to push this technology further to more species, strategies that communicate, and finally towards stochastic strategies on a lattice. However, new methods may yet be developed to deal with these challenging problems.

\section*{Conclusions}
While evolutionary games can be described succinctly in mathematical terms, they can only be solved exactly for the simplest of cases. Approximate methods exist to deal with finite population sizes and (very small) mutation rates, but the general case of stochastic and conditional strategies evolving at realistic mutation rates in finite populations (possibly on a grid or a network) can only be solved by explicitly simulating the population dynamics, and letting the agents ``play it out", so to speak. In such realistic settings we find that the outcome of a competition might depend on the particular parameters of the simulation, that then take on the role of ``environmental" parameters. For example, the fraction of the population that is replaced per update determines the average number of iterated games the agent plays. If agents communicate with each other (using strategies where the move is conditional on one or more previous plays), the mutation rate creates noise in the communication channel (because either of the strategies playing each other may suddenly change). Because the reliability of communication is essential in cooperation, both an increased mutation rate or a reduced population size (or an increased replacement rate) make cooperation less likely~\citep{Iliopoulosetal2010}. 

In evolutionary biology, we are used to considering evolutionary history as a sequence of changes that occurred on the background of the species that existed at the same time. While (for example) for a some period in time mammals and dinosaurs co-existed, people and dinosaurs did not, and we do not study how these two types would compete against each other. Within a single niche, organisms compete with the mutants that they share a fairly recent ancestor with, rather than with all possible genotypes that can be constructed within the genetic framework. Yet, in evolutionary game theory a majority of the existing work focuses on a set of strategies, and then considers the competition between all those strategies, at the same time. We have advocated here that while we should try to consider the full set of possible strategies when evolving them, we should not introduce all possible strategies all the time, but rather let the system find those strategies that ``play well together", so that truly robust mutational groups can be found. This, we advocate, will put a capital `E' back into EGT, an `E' that we feel has been missing.

We thus have to conclude that the limit where mathematical solutions are feasible represents rather unrealistic environments (such as infinite population size, vanishing mutation rate, perfectly mixed population), that cannot capture the complexity of stochastic and conditional play, as well as the dynamics of stable groups of strategies, the quasistrategies. However, mathematical solutions (even if they are only approximations) still play an essential role in evolutionary game theory, because they describe the limiting cases of agent-based methods, which must be explored in order to validate the simulations. In a well-developed simulation framework the limiting cases should be checked often so as to ensure that the dynamics are well understood. Properly used, agent-based simulations can go where mathematics cannot~\citep{Adami2012}, but they should be treated like computational experiments, whose significance must be judged in the light of judiciously designed controls.

\subsubsection*{Acknowledgements}This work was supported by the National Science Foundation's BEACON Institute for the Study of Evolution in Action under contract No. DBI-0939454. We acknowledge the support of Michigan State University's High Performance Computing Centre and the Institute for Cyber-Enabled Research (iCER). This work was also supported by resources provided by the Open Science Grid~\citep{Pordes2008}, which is supported by the National Science Foundation and the U.S. Department of Energy's Office of Science.


\section*{References}
\begin{thebibliography}{100}
\expandafter\ifx\csname url\endcsname\relax
  \def\url#1{\texttt{#1}}\fi
\expandafter\ifx\csname urlprefix\endcsname\relax\def\urlprefix{URL }\fi
\expandafter\ifx\csname href\endcsname\relax
  \def\href#1#2{#2} \def\path#1{#1}\fi

\bibitem{NeumannMorgenstern1944}
J.~von Neumann, O.~Morgenstern, Theory of Games and Economic Behavior,
  Princeton University Press, Princeton, NJ, 1944.

\bibitem{MaynardSmithPrice1973}
J.~{Maynard Smith}, G.~Price, The logic of animal conflict, Nature 246 (1973)
  15--18.

\bibitem{MaynardSmith1982}
J.~{Maynard Smith}, Evolution and the Theory of Games, Cambridge University
  Press, Cambridge, UK, 1982.

\bibitem{Axelrod1984}
R.~Axelrod, The Evolution of Cooperation, Basic Books, New York, NY, 1984.

\bibitem{Weibull1995}
J.~W. Weibull, Evolutionary Game Theory, MIT Press, Cambridge, MA, 1995.

\bibitem{Dugatkin1997}
L.~A. Dugatkin, Cooperation Among Animals: An Evolutionary Perspective,
  Princeton University Press, Princeton, NJ, 1997.

\bibitem{HofbauerSigmund1998}
J.~Hofbauer, K.~Sigmund, Evolutionary Games and Population Dynamics, Cambridge
  University Press, Cambridge, UK, 1998.

\bibitem{Nowak2006}
M.~Nowak, Evolutionary Dynamics, Harvard University Press, Cambridge, MA, 2006.

\bibitem{Nash1950}
J.~F. Nash, Equilibrium points in N-person games, Proceedings of the National
  Academy of Sciences 36 (1950) 48--49.

\bibitem{Zeeman1980}
E.~C. Zeeman, Population dynamics from game theory, in: Z.~Nitecki, C.~Robinson
  (Eds.), Global Theory of Dynamical Systems. Lecture Notes in Mathematics,
  Vol. 819, Springer, N.Y., 1980, pp. 471--497.

\bibitem{HofbauerSigmund2003}
J.~Hofbauer, K.~Sigmund, Evolutionary game dynamics, Bull. Am. Math. Soc. 40
  (2003) 479--519.

\bibitem{TaylorJonker1978}
P.~D. Taylor, L.~D. Jonker, Evolutionarily stable strategies and game dynamics,
  Mathematical Biosciences 40 (1978) 145--156.

\bibitem{SzaboFath2007}
G.~Szab\'o, G.~Fath, Evolutionary games on graphs, Physics Reports 446
  (2007) 97--216.

\bibitem{SzaboBorsos2016}
G.~Szab{\'o}, I.~Borsos, Evolutionary potential games on lattices, Physics
  Reports 624 (2016) 1--60.

\bibitem{Adamietal2012}
C.~Adami, J.~Schossau, A.~Hintze, Evolution and stability of altruist
  strategies in microbial games, Phys Rev E 85 (2012) 011914.

\bibitem{Price1970}
G.~R. Price, Selection and covariance, Nature 227 (1970) 520--521.

\bibitem{Fisher1930}
R.~Fisher, The Genetical Theory of Natural Selection, Oxford University Press,
  Oxford, UK, 1930.

\bibitem{Edwards1994}
A.~Edwards, The fundamental theorem of natural selection, Biological Reviews 69
  (1994) 443--474.

\bibitem{Holland2012}
J.~H. Holland, Signals and Boundaries: Building Blocks for Complex Adaptive
  Systems, MIT Press, Cambridge, MA, 2012.

\bibitem{Adami2012}
C.~Adami, Boldly going beyond mathematics, Science 338 (2012) 1421--1422.

\bibitem{ClaussenTraulsen2008}
J.~C. Claussen, A.~Traulsen, Cyclic dominance and biodiversity in well-mixed
  populations, Physical Review Letters 100 (2008) 058104.

\bibitem{TraulsenHauert2009}
A.~Traulsen, C.~Hauert, Stochastic evolutionary game dynamics, in: H.-G.
  Schuster (Ed.), Reviews of Nonlinear Dynamics and Complexity, Vol.~2,
  Wiley-VCH, Weinheim, 2009, pp. 25--63.

\bibitem{NowakSigmund1993}
M.~Nowak, K.~Sigmund, A strategy of win-stay, lose-shift that outperforms
  tit-for-tat in the prisoner's dilemma game, Nature 364 (1993) 56--58.

\bibitem{Eigen1971}
M.~Eigen, Selforganization of matter and the evolution of biological
  macromolecules, Die Naturwissenschaften 58 (1971) 465--523.

\bibitem{EigenSchuster1979}
M.~Eigen, P.~Schuster, The Hypercycle: A Principle of Natural
  Self-Organization, Springer-Verlag, Berlin, 1979.

\bibitem{SniegowskiGerrish2010}
P.~D. Sniegowski, P.~J. Gerrish, Beneficial mutations and the dynamics of
  adaptation in asexual populations, Philos Trans R Soc Lond B Biol Sci
  365 (2010) 1255--63.

\bibitem{Ewens2004}
W.~J. Ewens, Mathematical Population Genetics, 2nd Edition, Springer Science+
  Business Media, New York, NY, 2004.

\bibitem{DesaiFisher2007}
M.~M. Desai, D.~S. Fisher, Beneficial mutation selection balance and the effect
  of linkage on positive selection, Genetics 176 (2007) 1759--98.

\bibitem{Barricketal2009}
J.~E. Barrick, D.~S. Yu, S.~H. Yoon, H.~Jeong, T.~K. Oh, D.~Schneider, R.~E.
  Lenski, J.~F. Kim, Genome evolution and adaptation in a long-term experiment
  with {\it Escherichia coli}, Nature 461 (2009) 1243--1247.

\bibitem{GerrishLenski1998}
P.~J. Gerrish, R.~E. Lenski, The fate of competing beneficial mutations in an
  asexual population, Genetica 102-103 (1998) 127--144.

\bibitem{Maddamsettietal2015}
R.~Maddamsetti, R.~E. Lenski, J.~E. Barrick, Adaptation, clonal interference,
  and frequency-dependent interactions in a long-term evolution experiment with
  {\it Escherichia coli}, Genetics 200 (2015) 619--31.

\bibitem{Desaietal2007}
M.~M. Desai, D.~S. Fisher, A.~W. Murray, The speed of evolution and maintenance
  of variation in asexual populations, Curr Biol 17 (2007) 385--94.

\bibitem{Hines1980}
W.~G.~S. Hines, Strategy stability in complex populations, Journal of Applied
  Probability 17 (1980) 600--610.

\bibitem{Zeeman1981}
E.~C. Zeeman, Dynamics of the evolution of animal conflicts, Journal of
  Theoretical Biology 89 (1981) 249--270.

\bibitem{Thomas1985}
B.~Thomas, Evolutionarily stable sets in mixed-strategist models, Theoretical
  Population Biology 28 (1985) 332--341.

\bibitem{Bomze1990}
I.~M. Bomze, Dynamical aspects of evolutionary stability, Monatsh. Math. 110
  (1990) 189--206.

\bibitem{Cressman1990}
R.~Cressman, Strong stability and density-dependent evolutionarily stable
  strategies, J. Theoret. Biol. 145 (1990) 319--330.

\bibitem{LenskiVelicer2000}
R.~E. Lenski, G.~J. Velicer, Games microbes play, Selection 1 (2000) 51--57.

\bibitem{ChaoLevin1981}
L.~Chao, B.~R. Levin, Structured habitats and the evolution of anticompetitor
  toxins in bacteria, Proc Natl Acad Sci U S A 78~(10) (1981) 6324--8.

\bibitem{Kerretal2002}
B.~Kerr, M.~A. Riley, M.~W. Feldman, B.~J.~M. Bohannan, Local dispersal
  promotes biodiversity in a real-life game of rock-paper-scissors, Nature
  418 (2002) 171--4.

\bibitem{Thomas1985a}
B.~Thomas, On evolutionarily stable sets, Journal of Mathematical Biology 22
  (1985) 105---115.

\bibitem{Thomas1985b}
B.~Thomas, Evolutionary stable sets in mixed-strategist models, Theoretical
  Population Biology 28 (1985) 332--341.

\bibitem{Cressman2003}
R.~Cressman, Evolutionary Dynamics and Extensive Form Games, MIT Press,
  Cambridge, MA, 2003.

\bibitem{Michalewicz1996}
Z.~Michalewicz, Genetic Algorithms + Data Strucures = Evolution Programs,
  Springer Verlag, New York, 1996.

\bibitem{Iliopoulosetal2010}
D.~Iliopoulos, A.~Hintze, C.~Adami, Critical dynamics in the evolution of
  stochastic strategies for the iterated {P}risoner's {D}ilemma, PLoS
  Computational Biology 7 (2010) e1000948.

\bibitem{Moran1962}
P.~A.~P. Moran, The Statistical Processes of Evolutionary Theory, Clarendon
  Press, Oxford, 1962.

\bibitem{DonnellyWeber1985}
P.~Donnelly, N.~Weber, The {Wright-Fisher} model for temporally varying
  selection and population size, J. Math. Biol. 22 (1985) 21--29.

\bibitem{Lenskietal2003}
R.~E. Lenski, C.~Ofria, R.~T. Pennock, C.~Adami, The evolutionary origin of
  complex features., Nature 423 (2003) 139--144.

\bibitem{Ostmanetal2012}
B.~{\O}stman, A.~Hintze, C.~Adami, Impact of epistasis and pleiotropy on
  evolutionary adaptation, Proceedings of the Royal Society B 279 (2012)
  247--256.

\bibitem{Hilbeetal2013}
C.~Hilbe, M.~Nowak, K.~Sigmund, Evolution of extortion in iterated prisoner's
  dilemma games, Proc Natl Acad Sci USA 110 (2013) 6913--6918.

\bibitem{FletcherDoebeli2009}
J.~A. Fletcher, M.~Doebeli, A simple and general explanation for the evolution
  of altruism, Proc Biol Sci 276 (2009) 13--9.

\bibitem{Rapoportetal2015}
A.~Rapoport, D.~A. Seale, A.~M. Colman, Is tit-for-tat the answer? {On} the
  conclusions drawn from Axelrod's tournaments, PLoS One 10~(7) (2015)
  e0134128.

\bibitem{Nowak1990}
M.~Nowak, Stochastic strategies in the {Prisoner's Dilemma}, Theoretical
  Population Biology 38 (1990) 93--112.

\bibitem{NowakSigmund1990}
M.~Nowak, K.~Sigmund, The evolution of stochastic strategies in the {Prisoner's
  Dilemma}, Acta Appl. Math. 20 (1990) 247--265.

\bibitem{AdamiHintze2013}
C.~Adami, A.~Hintze, Evolutionary instability of zero-determinant strategies
  demonstrates that winning is not everything, Nature Communications 4 (2013)
  2193.

\bibitem{Akin2012}
E.~Akin, Stable cooperative solutions for the iterated {Prisoner's Dilemma}.,
  arxiv:1211.0969 (2012).

\bibitem{BoydLorberbaum1987}
R.~Boyd, J.~P. Lorberbaum, No pure strategy is evolutionarily stable in the
  repeated prisoner's dilemma game, Nature 327 (1987) 58--59.

\bibitem{Mirmomeni2015}
M.~Mirmomeni, Evolution of cooperation in the light of information theory,
  Ph.D. thesis, Michigan State University (2015).

\bibitem{PressDyson2012}
W.~Press, F.~J. Dyson, Iterated prisoners' dilemma contains strategies that
  dominate any evolutionary opponent, Proc Natl Acad Sci U S A 109 (2012) 10409--10413.

\bibitem{Boerlijstetal1997}
M.~C. Boerlijst, M.~A. Nowak, K.~Sigmund, Equal pay for all prisoners, Am.
  Math. Mon 104 (1997) 303--307.

\bibitem{StewartPlotkin2012}
A.~J. Stewart, J.~B. Plotkin, Extortion and cooperation in the prisoner's
  dilemma, Proc Natl Acad Sci U S A 109 (2012) 10134--5.

\bibitem{StewartPlotkin2013}
A.~Stewart, J.~Plotkin, From extortion to generosity, the evolution of
  zero--determinant strategies in the prisoner's dilemma, Proc Natl Acad Sci U
  S A 110 (2013) 15348--15353.

\bibitem{Leeetal2015}
C.~Lee, M.~Harper, D.~Fryer, The art of war: Beyond memory-one strategies in
  population games, PLoS One 10 (2015) e0120625.

\bibitem{AdamiHintze2016}
C.~Adami, A.~Hintze, Quasistrategies in stochastic evolutionary games,
  Forthcoming (2016).

\bibitem{SchusterSwetina1988}
P.~Schuster, J.~Swetina, Stationary mutant distributions and evolutionary
  optimization, Bull Math Biol 50~(6) (1988) 635--60.

\bibitem{Wilkeetal2001}
C.~O. Wilke, J.~L. Wang, C.~Ofria, R.~E. Lenski, C.~Adami, Evolution of digital
  organisms at high mutation rate leads to survival of the flattest, Nature 412
  (2001) 331--333.

\bibitem{WilkeAdami2003}
C.~O. Wilke, C.~Adami, Evolution of mutational robustness, Mutat Res 522
  (2003) 3--11.

\bibitem{Nimwegenetal1999}
E.~van Nimwegen, J.~P. Crutchfield, M.~Huynen, Neutral evolution of mutational
  robustness, Proc Natl Acad Sci U S A 96 (1999) 9716--20.

\bibitem{Wilke2001}
C.~O. Wilke, Adaptive evolution on neutral networks, Bulletin of Mathematical
  Biology 63 (2001) 715--730.

\bibitem{Wilke2005}
C.~O. Wilke, Quasispecies theory in the context of population genetics, BMC
  Evol Biol 5 (2005) 44.

\bibitem{Olson1971}
M.~Olson, The logic of collective action: Public goods and the theory of
  groups, Harvard University Press, Cambridge, MA, 1971.

\bibitem{DavisHolt1993}
D.~D. Davis, C.~A. Holt, Experimental Economics, Princeton University Press,
  Princeton, N.J., 1993.

\bibitem{Ledyard1995}
J.~Ledyard, Public goods: A survey of experimental research, in: J.~H. Kagel,
  A.~E. Roth (Eds.), Handbook of experimental economics, Princeton University
  Press, Princeton, N.J., 1995, pp. 111--194.

\bibitem{Hardin1968}
G.~Hardin, The tragedy of the commons, Science 162 (1968) 1243--1248.

\bibitem{HintzeAdami2015}
A.~Hintze, C.~Adami, Punishment in public goods games leads to meta-stable
  phase transitions and hysteresis, Phys Biol 12 (2015) 046005.

\bibitem{ImryWortis1979}
Y.~Imry, M.~Wortis, Influence of quenched impurities on first-order phase
  transitions, Phys. Rev. B (1979) 3580--3585.

\bibitem{Hilbeetal2014}
C.~Hilbe, B.~Wu, A.~Traulsen, M.~A. Nowak, Cooperation and control in
  multiplayer social dilemmas, Proc Natl Acad Sci U S A 111 (2014)
  16425--30.

\bibitem{Hilbeetal2015}
C.~Hilbe, B.~Wu, A.~Traulsen, M.~A. Nowak, Evolutionary performance of
  zero-determinant strategies in multiplayer games, J Theor Biol 374 (2015)
  115--24.

\bibitem{Panetal2015}
L.~Pan, D.~Hao, Z.~Rong, T.~Zhou, Zero-determinant strategies in iterated
  public goods game, Sci Rep 5 (2015) 13096.

\bibitem{HauertSchuster1997}
C.~Hauert, H.~G. Schuster, Effects of increasing the number of players and
  memory size in the iterated prisoner's dilemma: A numerical approach, Proc.
  Roy. Soc. London B 264 (1997) 513--519.

\bibitem{Yamagishi1986}
T.~Yamagishi, The provision of a sanctioning system as a public good, Journal
  of Personality and Social Psychology 51 (1986) 110--116.

\bibitem{Sigmundetal2001}
K.~Sigmund, C.~Hauert, M.~A. Nowak, Reward and punishment, Proc Natl Acad Sci U
  S A 98 (2001) 10757--62.

\bibitem{SzaboHauert2002}
G.~Szabo, C.~Hauert, Phase transitions and volunteering in spatial public goods
  games, Physical Review Letters 89 (2002) 118101.

\bibitem{FehrGachter2002}
E.~Fehr, S.~Gachter, Altruistic punishment in humans, Nature 415 (2002)
  137--140.

\bibitem{FehrFischbacher2003}
E.~Fehr, U.~Fischbacher, The nature of human altruism, Nature 425 (2003)
  785--791.

\bibitem{Hammerstein2003}
P.~Hammerstein (Ed.), Genetic and Cultural Evolution of Cooperation, MIT Press,
  Cambridge, MA, 2003.

\bibitem{Fowler2005}
J.~H. Fowler, Altruistic punishment and the origin of cooperation, Proc Natl
  Acad Sci U S A 102 (2005) 7047--9.

\bibitem{NakamaruIwasa2006}
M.~Nakamaru, Y.~Iwasa, The coevolution of altruism and punishment: Role of the
  selfish punisher, J Theor Biol 240 (2006) 475--88.

\bibitem{CamererFehr2006}
C.~F. Camerer, E.~Fehr, When does ``economic man" dominate social behavior?,
  Science 311 (2006) 47--52.

\bibitem{Gurerketal2006}
O.~G{\"u}rerk, B.~Irlenbusch, B.~Rockenbach, The competitive advantage of
  sanctioning institutions, Science 312 (2006) 108--11.

\bibitem{Hauertetal2007}
C.~Hauert, A.~Traulsen, H.~Brandt, M.~A. Nowak, K.~Sigmund, Via freedom to
  coercion: The emergence of costly punishment, Science 316 (2007)
  1905--7.

\bibitem{DeSilvaetal2009}
H.~D. Silva, C.~Hauert, A.~Traulsen, K.~Sigmund, Freedom, enforcement, and the
  social dilemma of strong altruism, J. Evol. Econ. 20 (2009) 203--217.

\bibitem{HenrichBoyd2001}
J.~Henrich, R.~Boyd, Why people punish defectors. weak conformist transmission
  can stabilize costly enforcement of norms in cooperative dilemmas, J Theor
  Biol 208 (2001) 79--89.

\bibitem{Boydetal2003}
R.~Boyd, H.~Gintis, S.~Bowles, P.~J. Richerson, The evolution of altruistic
  punishment, Proc Natl Acad Sci U S A 100 (2003) 3531--5.

\bibitem{Brandtetal2003}
H.~Brandt, C.~Hauert, K.~Sigmund, Punishment and reputation in spatial public
  goods games, Proc Biol Sci 270~(1519) (2003) 1099--104.

\bibitem{Helbingetal2010}
D.~Helbing, A.~Szolnoki, M.~Perc, G.~Szab\'o, Evolutionary establishment of
  moral and double moral standards through spatial interactions, PLoS Comput
  Biol 6 (2010) e1000758.

\bibitem{Helbingetal2010c}
D.~Helbing, A.~Szolnoki, M.~Perc, G.~Szab\'o, Punish, but not too hard: How
  costly punishment spreads in the spatial public goods game, New Journal of
  Physics 12 (2010) 083005.

\bibitem{Boydetal2010}
R.~Boyd, H.~Gintis, S.~Bowles, Coordinated punishment of defectors sustains
  cooperation and can proliferate when rare, Science 328 (2010) 617--20.

\bibitem{Sigmundetal2010}
K.~Sigmund, H.~De~Silva, A.~Traulsen, C.~Hauert, Social learning promotes
  institutions for governing the commons, Nature 466 (2010) 861--3.

\bibitem{Sasakietal2011}
T.~Sasaki, T.~Unemi, Replicator dynamics in public goods games with reward
  funds, J Theor Biol 287 (2011) 109--14.

\bibitem{Szolnokietal2011}
A.~Szolnoki, G.~Szab{\'o}, M.~Perc, Phase diagrams for the spatial public goods
  game with pool punishment, Phys Rev E 83~(3 Pt 2) (2011) 036101.

\bibitem{PercSzolnoki2012}
M.~Perc, A.~Szolnoki, Self-organization of punishment in structured
  populations, New Journal of Physics 14 (2012) 043013.

\bibitem{SzolnokiPerc2013a}
A.~Szolnoki, M.~Perc, Effectiveness of conditional punishment for the evolution
  of public cooperation, J Theor Biol 325 (2013) 34--41.

\bibitem{SzolnokiPerc2013b}
A.~Szolnoki, M.~Perc, Correlation of positive and negative reciprocity fails to
  confer an evolutionary advantage: Phase transitions to elementary strategies,
  Phys Rev X 3 (2013) 041021.

\bibitem{Chenetal2014}
X.~Chen, A.~Szolnoki, M.~Perc, Probabilistic sharing solves the problem of
  costly punishment, New Journal of Physics 16 (2014) 083016.

\bibitem{LandauLifshitz1969}
L.~D. Landau, E.~M. Lifshitz, Statistical Physics (Course of Theoretical
  Physics Part 5), 2nd Edition, Pergamon Press, Oxford (UK), 1969.

\bibitem{Adamietal2015}
C.~Adami, N.~Pasmanter, A.~Hintze, Thermodynamics of evolutionary games, In
  preparation (2016).

\bibitem{NowakMay1992}
M.~A. Nowak, R.~M. May, Evolutionary games and spatial chaos, Nature 359 (1992)
  826--829.

\bibitem{NowakMay1993}
M.~A. Nowak, R.~M. May, The spatial dilemmas of evolution, Int J Bifurcat Chaos
  3 (1993) 35--78.

\bibitem{Nowaketal1994}
M.~A. Nowak, S.~Bonhoeffer, R.~M. May, Spatial games and the maintenance of
  cooperation, Proc Natl Acad Sci U S A 91 (1994) 4877--4881.

\bibitem{HauertDoebeli2004}
C.~Hauert, M.~Doebeli, Spatial structure often inhibits the evolution of
  cooperation in the snowdrift game., Nature 428 (2004) 643--646.

\bibitem{SzaboToke1998}
G.~Szabo, C.~Toke, Evolutionary prisoner's dilemma game on a square lattice,
  Phys Rev E 58 (1998) 69--73.

\bibitem{Szolnokietal2009}
A.~Szolnoki, M.~Perc, G.~Szabo, Phase diagrams for three-strategy evolutionary
  prisoner's dilemma games on regular graphs., Phys Rev E 80~(5 Pt 2) (2009)
  056104.

\bibitem{Wolfram1984}
S.~Wolfram, Cellular automata as models of complexity, Nature 311 (1984)
  419--424.

\bibitem{Reichenbachetal2007}
T.~Reichenbach, M.~Mobilia, E.~Frey, Mobility promotes and jeopardizes
  biodiversity in rock-paper-scissors games, Nature 448 (2007) 1046--9.

\bibitem{Pordes2008}
{R. Pordes et al}., The open science grid, J. Phys. Conf. Ser. 78 (2007)
  012057.

\end{thebibliography}
\end{document}